\documentclass[aps,preprint,nofootinbib,preprintnumbers,eqsecnum]{revtex4-1}
\usepackage[usenames, dvipsnames]{color}

\usepackage{graphicx}
\usepackage{xcolor}
\usepackage{rotating}
\usepackage{bm,amsmath,amssymb}
\usepackage[utf8]{inputenc}
\usepackage{footmisc}

\newcommand{\nb}[1]{\color{blue}}

\newcommand{\HL}[1]{{\textcolor{magenta}{#1}}}
\newcommand{\ML}[1]{{\textcolor[rgb]{.2,.6,0.5}{#1}}}
\newcommand{\hl}[1]{\color{magenta}}
\newcommand{\para}{{\parallel}}

\usepackage[
	  pagebackref=false,
	  colorlinks=true,
      linkcolor=blue,
      urlcolor=blue,
      filecolor=black,
      citecolor=red,
      pdfstartview=FitV,
      pdftitle={},
        pdfauthor={},
        pdfsubject={},
        pdfkeywords={},
        pdfpagemode=None,
        bookmarksopen=true
      ]{hyperref}

\usepackage{graphicx}
\graphicspath{ {./images/} }

\usepackage{caption}
\usepackage{subcaption}

\usepackage{comment}

\usepackage[normalem]{ulem}
\usepackage{amsmath}
\usepackage{enumerate}
\usepackage{amsfonts}
\usepackage{epsfig}
\usepackage{mathbbol}

\newcommand\p{\ensuremath{\partial}}
\newcommand\pp{\ensuremath{\vec\nabla}}

\newcommand{\be}{\begin{equation}}
\newcommand{\ee}{\end{equation}}
\newcommand{\bea}{\begin{eqnarray}}
\newcommand{\eea}{\end{eqnarray}}
\newcommand{\bega}{\begin{gather}}
\newcommand{\eega}{\end{gather}}

\newcommand{\bi}{\begin{itemize}}
\newcommand{\ei}{\end{itemize}}
\newcommand{\ben}{\begin{enumerate}}
\newcommand{\een}{\end{enumerate}}
\newcommand{\bca}{\begin{cases}}
\newcommand{\eca}{\end{cases}}
\newcommand{\bln}{\begin{align}}
\newcommand{\eln}{\end{align}}
\newcommand{\bst}{\begin{split}}
\newcommand{\est}{\end{split}}
\def\ie{\begin{equation}\begin{aligned}}
\def\fe{\end{aligned}\end{equation}}
\newcommand{\bma}{\le(\begin{matrix}}
\newcommand{\ema}{\end{matrix}\ri)}
\newcommand{\bwt}{\begin{widetext}}
\newcommand{\ewt}{\end{widetext}}

\newcommand\al{{\alpha}}
\def\b{{\beta}}
\newcommand\ep{\epsilon}
\newcommand\sig{\sigma}

\newcommand\lam{\lambda}
\newcommand\Lam{\Lambda}
\newcommand\om{\omega}

\newcommand\ga{{\ensuremath{{\gamma}}}}
\newcommand\Ga{{\ensuremath{{\Gamma}}}}
\newcommand\de{{\ensuremath{{\delta}}}}
\newcommand\De{{\ensuremath{{\Delta}}}}
\newcommand\vp{\varphi}
\newcommand\ka{\kappa}

\newcommand\nab{{\nabla}}
\newcommand\Th{{\Theta}}

\newcommand\ov{\over}

\def\le{\left}
\def\ri{\right}

\newcommand\sA{{\ensuremath{{\mathcal A}}}}
\newcommand\sB{{\ensuremath{{\mathcal B}}}}
\newcommand\sC{{\ensuremath{{\mathcal C}}}}
\newcommand\sD{{\ensuremath{{\mathcal D}}}}
\newcommand\sE{{\ensuremath{{\mathcal E}}}}

\newcommand\sG{{\ensuremath{{\mathcal G}}}}
\newcommand\sH{{\ensuremath{{\mathcal H}}}}

\newcommand\sL{{\ensuremath{{\mathcal L}}}}

\newcommand\sP{{\ensuremath{{\mathcal P}}}}

\newcommand\sT{{\mathcal T}}

\newcommand\vx{{\vec x}}
\newcommand\vk{{\vec k}}

\DeclareMathAlphabet{\pazocal}{OMS}{zplm}{m}{n}

\newcommand{\cL}{{\pazocal L}}

\newcommand{\dcmw}{{chiral magnetic electric separation wave}}
\newcommand{\Dcmw}{{Chiral magnetic electric separation wave}}

\begin{document}

\preprint{MIT-CTP/5509}

\title{A systematic formulation of chiral anomalous magnetohydrodynamics}

\author{Michael Landry and Hong Liu}
\affiliation{Center for Theoretical Physics, MIT, Cambridge, MA 02139, USA 
}

\begin{abstract}

We present a new way of deriving effective theories of dynamical electromagnetic fields in general media. 
It can be used to give a systematic formulation of magnetohydrodynamics (MHD) with strong magnetic fields, including 
 systems with chiral matter and Adler-Bell-Jackiw (ABJ) anomaly. We work in the regime in which velocity and temperature fluctuations can be neglected. 
The resulting chiral anomalous  MHD incorporates and generalizes the chiral magnetic effect, the chiral separation effect, the chiral electric separation effect,
as well as recently derived strong-field MHD, all in a single coherent framework. 
At linearized level, the theory predicts that the
chiral magnetic wave does not survive dynamical electromagnetic fields. A different chiral wave, to which we refer as the \dcmw, emerges 
as a result of dynamical versions of the chiral electric separation effect and the chiral magnetic effect.
We predict its wave velocity. We also introduce a simple, but solvable nonlinear model to explore the fate of the chiral instability. 

\end{abstract}
\maketitle
\tableofcontents

\section{Introduction} 


Consider a system in which electromagnetic fields interact strongly with charged matter at a finite temperature. 
The magnetohydrodynamic approximation applies to the regime in which, in addition to conserved quantities, the only other ``slow'' variables are magnetic fields\footnote{In fact, from a modern perspective~\cite{Gaiotto:2014kfa,Grozdanov:2016tdf}, magnetic fields can also be viewed as conserved charges associated with a one-form symmetry.}; MHD is a universal effective theory for these slow variables. 
MHD is a powerful framework with applications to a wide range of disciplines  including nuclear physics, condensed matter physics, astrophysics, and cosmology. In systems with chiral matter, the chiral symmetry can be broken by the Adler-Bell-Jackiw (ABJ) anomaly, in which case the chiral charge density is not conserved, and strictly speaking drops out of the hydrodynamic description.  
However, if the anomaly coefficient is sufficiently small such that the time and length scales responsible for the non-conservation of the chiral symmetry is much larger than typical microscopic relaxation scales, we can treat the chiral charge density as an approximately conserved quantity and keep it in the formulation of MHD, resulting in a chiral anomalous MHD. 

There is a large literature pertaining to macroscopic effects of quantum anomalies on transports; with effects from t' Hooft anomalies in which the external fields are not dynamical, these effects are fairly well understood (see~\cite{Landsteiner:2016led,Kharzeev:2015znc} for reviews). 
For ABJ anomalies with dynamical electromagnetic fields, the story is more intricate, see e.g.~\cite{Golkar:2012kb,Hou:2012xg,Jensen:2013vta,Gorbar:2013upa,Hirono:2015rla,Figueroa:2017hun,Figueroa:2019jsi,Das:2022auy,Nair:2011mk,Capasso:2013jva,Monteiro:2014wsa} for discussions. There have also been many discussions of chiral anomalous  MHD in the weak magnetic field regime~\cite{Boyarsky:2015faa,Pavlovic:2016gac,Giovannini:2016whv,Yamamoto:2016xtu,Rogachevskii:2017uyc,Hattori:2017usa,Hattori:2022hyo}.
In this paper we present a systematic formulation of chiral anomalous MHD in the strong field regime. We first present a formulation of EFTs for electromagnetic fields in general media, and then specify to MHD. The incorporation of the ABJ anomaly can be readily done in this approach. We will outline the main idea and results, leaving technical details to a longer exposition~\cite{LandryLiu}. 


{\bf Note added:}
While this paper was nearing completion, we learned of~\cite{Iqbaletal}, which approaches chiral magnetohydrodynamics 
using one-form symmetry. 


\section{EFT of dynamical electromagnetism in general media}

Consider the electromagnetic vector field $a_\mu$ in a general medium at a finite temperature $T_0= 1/\b_0$. 
We will work with the normal phase, i.e. the $U(1)$ gauge symmetry is not spontaneously broken. 
For the moment, let us assume that other than the $U(1)$ gauge symmetry, the system does not have any other internal continuous global symmetry. 
We are interested in the behavior of 
the electromagnetic field at distance and time scales much larger than the typical relaxation scale, which we will denote as $\ell$.\footnote{For notational simplicity, we use $\ell$ to also denote the relaxation time assuming there is a typical velocity to convert one to the other.}
For this purpose, we can imagine integrating out all  matter fields to obtain an effective action for $a_\mu$. 
We will ignore the motion of the medium and temperature fluctuations, although they can be readily taken into account using the method discussed below. Such a generalization will be presented elsewhere. 
To formulate a real-time action we should use the Schwinger-Keldysh (or closed time path) formalism~\cite{Keldysh:1964ud} in which we have two copies of $a_\mu$, written in terms of the so-called $r-a$ variables $a_{r \mu }$ and $a_{a \mu }$ (below we will suppress $r$ in $a_{r \mu}$), with $a_\mu$  the physical field and $a_{a \mu}$ the corresponding noise field.


The effective action $I_{\rm EFT}$ for the dynamical electromagnetic field can be obtained using the framework developed in~\cite{Crossley:2015evo,Glorioso:2017fpd,Glorioso:2018wxw} as follows. 
We first imagine the $U(1)$ symmetry is global, and $a_\mu, a_{a \mu}$ are external sources for two copies of the conserved current. 
The hydrodynamical variables for the current conservation are two scalars $\phi, \phi_a$. 
The hydrodynamic action $I_{\rm hydro}$ depends on 
$a_\mu, a_{a \mu}, \phi, \phi_a$ only through the combinations $A_\mu = a_\mu + \p_\mu \phi$ and $A_{\mu a} = a_{a\mu} + \p_\mu \phi_a$, i.e. $I_{\rm hydro}  =I_{\rm hydro} [A_\mu , A_{\mu a}]$, and satisfies the following conditions:

\ben 

\item It enjoys translational invariance in $x^\mu = (t, x^i) $ and rotationally invariance in $x^i$ directions. 

\item There are various unitarity constraints (suppressing $\mu$ indices): 
\bega  \label{fer1}
 I^*_{\rm EFT} [A, A_a]  = - I_{\rm EFT} [A, - A_a], 
\\
 \label{pos}
{\rm Im} \, I_{\rm EFT} \geq 0, \\
\label{1keyp2}
I_{\rm EFT}[A,  A_a = 0 ]=0\ .
\end{gather}

\item It is invariant under the time-independent local transformation 
\be \label{daug}
 A_{i} \to  A_{i}' = A_{i} - \p_i \lam (x^i) \ ,
 \ee
 which means that the action can depend on $A_i$ only through $\p_t A_i$ or $\vec B = \nab \times \vec A$.\footnote{$\vec A$ denotes the spatial vector.} 

\item  It is invariant under dynamical KMS transformations, i.e. $I_{\rm EFT}  [A, A_a] = I_{\rm EFT}  [\tilde A , \tilde A_a]$ where  
dynamical KMS transformations are $Z_2$ transformations of the form 
 \be \label{kms}
\tilde A_{\mu} (x) = \Th A_{\mu}  (x) , \quad \tilde A_{\mu a } (x) = \Th A_{a \mu}  (x) + i \b_0  \Th \p_t  A_{\mu} (x)  \ .
\ee
Here $\Th$ is a discrete symmetry of the underlying microscopic system involving time reversal. 
The dynamical KMS symmetry as well as unitarity constraints~\eqref{fer1}--\eqref{1keyp2} ensure that the theory has an entropy current with non-negative divergence~\cite{Glorioso:2016gsa}. 
 
 \item Impose whatever other discrete symmetries (formed from $\sC, \sP$) preserved by the underlying microscopic system.\footnote{If a system has some discrete internal symmetries, they should of course also be imposed.} 

\een

Now promote $a_\mu, a_{a \mu}$ to be dynamical fields (i.e. gauging the $U(1)$ symmetry) by integrating over them in the path integrals. 
We can then absorb  $\phi, \phi_a$  into the redefinitions of them to obtain 
\be 
\int D a_\mu D a_{a \mu} D \phi D \phi_a \, e^{i I_{\rm hydro}  [A_\mu, A_{a \mu}]} 
=  C \int D A_\mu D A_{a \mu} e^{i I_{\rm EFT}  [A_\mu, A_{a \mu}]} 
\ee
where $\phi, \phi_a$ integrals drop out, $C$ is an infinite constant, and 
\be 
I_{\rm EFT} [A_\mu, A_{a \mu}] = I_{\rm hydro} [A_\mu , A_{a\mu} ] \ .
\ee
$A_\mu, A_{a\mu}$ are now dynamical fields which give the same field strength as $a_\mu, a_{a\mu}$. 
They are invariant under ``gauge'' transformations of $a_{\mu}, a_{a\mu}$ as any such transformations can be absorbed by shifts of $\phi, \phi_a$, in analogue with the standard Higgs mechanism. 
But because of~\eqref{daug}, $A_i$ still transforms under a local symmetry. While~\eqref{daug}, referred to as the diagonal shift in~\cite{Crossley:2015evo}, has nothing to do with the original gauge symmetry, it is as if there is some residual gauge symmetry left. 
With a slight abuse of language, we will say that $a_\mu$ undergoes a  ``partial Higgs mechanism'' to become $A_\mu$.\footnote{In a superconducting phase, where the $U(1)$ gauge symmetry is spontaneously broken, we drop the requirement to be invariant under~\eqref{daug}.}

To formulate an action we also need to choose a derivative expansion scheme. There are three possible regimes. The first is the standard dielectric regime where $A_\mu$ is assigned weight $0$ and $\p_t, \p_i$ weight $1$, {which applies to the regime of ``weak'' interactions of electromagnectic field with insulators.}
The second is the MHD regime where $A_i$ is assigned  weight $-1$, $A_0$ weight $0$, $\p_i$ weight $1$,  $\p_t$ weight $2$. Thus $B_i$ has weight zero and $E_i$ has weight $1$. This weight assignment is motivated by the fact that in the MHD regime the dominant behavior is magnetic diffusion (i.e. $\p_t \sim \p_i^2$) and we would like to have arbitrary dependence on magnetic field $B_i$, which then must have weight $0$. The weight assignments for $A_{a\mu }$ can be deduced from those of $A_\mu$ by using~\eqref{kms}. The last possible scheme is for $A_\mu$ to have weight $-1$, and $\p_\mu$ weight $1$, which leads to both $E_i$ and $B_i$ having weight zero.\footnote{See~\cite{Kovtun:2016lfw} for a recent discussion of thermodynamics of such systems.}
In this paper we focus on the MHD regime, {leaving the others to~\cite{LandryLiu}.}

\section{MHD} 

As an example, suppose the only preserved discrete symmetry of a system is $\sT$, we then need to take $\Th = \sT$. 
In the MHD regime, up to weight 2 the Lagrangian density  has the form 
 \bega \label{ehb1} 
 \sL_{\sT} =  b_0 B_i A_{ai}   + n  A_{a0}  -  H_i B_{ai} + i  (r^{-1})_{ij}  A_{ai} 
 (T_0 A_{aj} + i \p_t A_j )
 \end{gather}
 where  $n$ and $H$ can be obtained from a single function $F (A_0, B^2)$ as 
 \be 
 n = -{\p F \ov \p A_0},  \quad H_i =  {\p F \ov \p B_i} \equiv {B_i \ov \mu}  , 
 \quad B^2 = \vec B^2 , \quad
h_i \equiv {B_i \ov B} 
 \ee
  and $r^{-1}$ is the inverse of matrix $r_{ij}$ whose matrix elements can be parameterized using rotational symmetry as 
 \be \label{ueh}
r_{ij} = r_\parallel h_i h_j + r_\perp \Delta_{ij} +r_\times \epsilon_{ijk} h_k,\quad      \Delta_{ij} = \delta_{ij} - h_i h_j \ . 
 \ee
 All coefficients in $r_{ij}$ are functions of $A_0$ and $B^2$, while $b_0$ is a constant. 
We may interpret $F$ as the equilibrium free energy density, and $A_0$ as the (dynamical) chemical potential.
Then $n$ can be interpreted as the local charge density, and $H_i$ is the 
medium magnetic field. From rotational symmetry $H_i$ must be proportional to $B_i$, and $\mu (A_0, B^2)$, can be interpreted as the 
magnetic permeability. 
Equation~\eqref{pos} implies that 
\be
r_{\parallel}, r_\perp \geq 0, 
\ee
 while $r_\times$ can have either sign.

 The equation of motion for $A_{a0}$ implies that $n =0$, i.e. the local charge density vanishes, which in turn implies  $A_0 = f (B^2)$, where $f$ is some function.
  Then all other coefficients $\mu, r_\parallel, r_\perp, r_\times$ can be viewed as functions of $B^2$. 
  With $b_0$ being a constant independent of any dynamical variables\footnote{When we include dynamical temperature, then $b_0$ will be replaced by ${b_0 \ov \b}$ with $b_0$ a constant. The term is still topological, as other than a universal $1/\b$, there is no dependence on physical condition of the system.}, the $b_0$ term should be viewed as a topological term. 
Thermodynamic stability also requires $\mu \geq 0$.

Note that nonzero $A_0 = f(B^2)$ and $r_\times$ are consequences of charge conjugation violation, while a nonzero $b_0$ is a consequence parity violation.\footnote{At the level of the action, we can have $r_\times\neq 0$ appear even when $\sC$ symmetry is respected; the requirement is merely that it be an odd function of $A_0$. But then $A_0=0$ on shell if $\sC$ symmetry is respected, so we ultimately find that $r_\times=0$ on the equations of motion.} If we consider instead a system where the only preserved discrete symmetry is $\Th = \sC \sP \sT$, then we find that
all of them vanish.

The equations of motion for $A_{ai}$ give
 \bega \label{teg}
E_i = \p_i A_0 - \p_0 A_i =  r_{ij} (\ep_{jkl} \p_k H_l- b_0 B_j ) + \p_i f,
\end{gather} 
which is the constitutive relation of the electric field in terms of the magnetic field. In the weak magnetic field limit, we can expand $r_\parallel (B^2=0) = r_\perp (B^2=0)  = r$. 
Equation~\eqref{teg} can be written to linear order in $B$ as 
\be 
\vec E = {r \ov \mu}  \nab \times \vec B + b_0 r \vec B , \quad \vec B \to 0 
\ee
which then identifies $r = {1 \ov \sig}$ with $\sig$  the electrical conductivity. {The second term proportion to $\vec B$ is reminiscent of a similar term in~\cite{PhysRevD.41.1231}.}
We stress that for general $\vec B$, $r_\parallel, r_\perp, r_\times$ are new transport coefficients and are not related to conductivity.

 MHD has recently been reformulated using the one-form symmetry~\cite{Grozdanov:2016tdf} (see~\cite{Hernandez:2017mch,Grozdanov:2017kyl,Glorioso:2018kcp,Armas:2018zbe,Gralla:2018kif,Benenowski:2019ule,Landry:2021kko,Armas:2022wvb,Das:2022auy,Vardhan:2022wxz} for various developments). In particular, the constitutive relation of the electric field that violates charge and parity conjugation in the strong magnetic field regime has recently been written down in~\cite{Vardhan:2022wxz}, generalizing the previous formulation of~\cite{1992ApJ...395..250G}. It can be checked that~\eqref{teg} is exactly equivalent to the expression in~\cite{Vardhan:2022wxz} except that the $b_0$ term was missing there. 
 See Appendix~\ref{app:b}. 
 It can be shown that the two formulations are related by a duality transformation, and the reason that the $b_0$ term is absent in~\cite{Vardhan:2022wxz}  is due to  the fact that the term becomes non-local in the one-form symmetry formulation. See~\cite{LandryLiu} for details.

The same approach can be applied to gravity to derive effective field theories of dynamical gravity strongly interacting with matter fields, and will be pursued elsewhere.  

\section{Chiral anomalous MHD} 

We now generalize the above discussion to including chiral matter, in which case there is a chiral current $J_\mu^5$
satisfying the equation 
\begin{equation}\label{an5}
 \partial_\mu J_5^{\mu} = \frac{c}{4} \epsilon^{\mu\nu\alpha\beta} F_{\mu\nu} F_{\alpha\beta} \ .
\end{equation}
where $F=dA$. As mentioned at the beginning, we will assume the anomaly coefficient $c$ is sufficiently small that the non-conservation due to~\eqref{an5} happens at much longer scales than the microscopic relaxation scale. 


Denote $\sA_\mu, \sA_{a\mu }$ and $\vp, \vp_a$ respectively as external sources and hydrodynamic degrees of freedom for $J^\mu_5$. We again suppressed subscript $r$ in $\sA_{\mu}$ and $\vp$. The EFT action should have the structure~\cite{Glorioso:2017lcn}
 \begin{equation}
I_\text{EFT} = I_\text{anom} + I_\text{inv},
\end{equation}
where $ I_\text{inv} =  I_\text{inv}  [A_{\mu}, A_{a \mu} ; C_\mu, C_{a \mu}]$, depending on $\vp, \vp_a$ only through the combinations
\be 
C_\mu = \sA_\mu + \p_\mu \vp, \quad
C_{\mu a} = \sA_{\mu a} + \p_\mu \vp_a , 
\ee
and $I_\text{anom}$ has the form~\cite{Glorioso:2017lcn}
\bea\label{I anom} 
 I_\text{anom} & = &  c  \int  (\varphi_a F\wedge F + 2\varphi F\wedge F_a ) \ .
\label{I anom ra}
  \eea
$I_\text{anom}$ is designed to capture the anomaly~\eqref{an5}, and indeed it can be readily checked that 
  equations of motion for $\vp, \vp_a$ give two copies of~\eqref{an5}, one for each leg of the S-K contour. 
 In addition to the conditions mentioned earlier, $I_{\rm EFT}$ should now also be invariant under 
the shift~\cite{Glorioso:2017lcn,Glorioso:2018wxw}
\begin{equation} \label{ds5}
 \varphi \to\varphi +\Lambda(\vec x) ,
\end{equation}
and the dynamical KMS transformation should now also include the transformation of $C_\mu, C_{a \mu}$
\be \label{kms5}
\tilde C_{\mu} (x) = \Th C_{\mu}  (x) , \quad \tilde C_{a\mu  } (x) = \Th C_{a \mu}  (x) + i \b_0  \Th \p_t  C_{\mu} (x)  \ .
\ee
Under $\sC, \sP, \sT$ transformations, $C_\mu, C_{a \mu}$ transform the same as $J_5^\mu$, i.e. 
\be 
\sC: C_\mu \to  C_\mu, \quad \sP: C_\mu \to (- C_0, C_i),  \quad 
  \sT: C_\mu  \to ( C_0, - C_i) \ .
\ee
The weight assignments are that $C_0 = \sA_0 + \p_t \vp \equiv \mu_5$ has weight $0$, while $\vec \sA, C_{a0}, \vec C_a$ respectively have weight $-1, 2, 1$. Here $\mu_5$ is the local chemical potential for the chiral charge density. 
Note that using the one-form symmetry formulation to describe the anomaly equation~\eqref{an5} is challenging as the right hand side is then expressed in terms of the two-form current which itself has to be determined dynamically. 


In the MHD regime the Lagrangian density up to weight $2$ has the form (to linear order in external sources $\sA_{a\mu}$)~\cite{LandryLiu}
\be\begin{split} \label{probe}
     \cL = n  A_{a0} + n_5 \p_t \varphi_a -H_i B_{ai} +2 c \partial_t \vec A \cdot\vec B \varphi_a -(2 c  \mu_5 -b_{00}) B_i A_{ai}  \\ 
     +i (r^{-1})_{ij} A_{ai}(T_0 A_{aj} + i \p_t A_j ) + i \kappa_{ij} \partial_i \varphi_a (T_0 \partial_j \varphi_a +i\partial_j \mu_5) \\
   +i  \lam_{ij}  (  (T_0 \partial_i \varphi_a +i\partial_i \mu_5) A_{aj}+ \p_i \vp_a  (T_0 A_{aj} + i \p_t A_j )) \\
+ \gamma_{ij} ( \partial_i \varphi_a \partial_t A_j -  \partial_i \mu_5 A_{aj})+ \sA_{a0} J^0_5 +  \sA_{a i} J^i_5 
\end{split}\ee
where $r_{ij}$ has the same decomposition as~\eqref{ueh}, and 
\bega
\label{desc}
n =- {\p F \ov \p A_0} , \quad n_5 = - \frac{\p F}{\p \mu_5},\quad H_i =    {\p F \ov \p B_i} \equiv {B_i \ov \mu},  
  \\
     \kappa_{ij} = \kappa_\parallel h_i h_j + \kappa_\perp \Delta_{ij} +\kappa_\times \epsilon_{ijk} h_k,\quad \lam_{ij} = \lam_\parallel h_i h_j  + \lam_\perp \Delta_{ij} +\lam_\times \epsilon_{ijk} h_k , 
     \\ \gamma_{ij} = \gamma_\parallel h_i h_j + \gamma_\perp \Delta_{ij} + \gamma_\times \epsilon_{ijk} h_k,\quad \lambda_{ij}^\pm \equiv \lam_{ij} \pm \gamma_{ij} \\
     J^0_5 =n_5, \quad 
     J^{i}_5 = (b_{50} -2 c A_0) B_i  - \ka_{ij} \p_j \mu_5   - \lam^-_{ij} \p_t A_j + \epsilon_{ijk} \partial_j(m B_k) .
     \label{yeq}
\end{gather} 
All coefficients should be understood as functions of $A_0, \mu_5, B^2$ except that $b_{00}, b_{50}$ are constants. 
Depending on the choice of $\Th$, some coefficients may vanish or have constraints. For example, 
for $\Th = \sT$, we have $\lambda_\times=\gamma_{\parallel,\perp}=0$. For $\Theta=\sC\sP\sT$, we have that $b_{00}=b_{50}=0$ and the coefficients $ F, r_{\parallel,\perp}, \kappa_{\parallel,\perp}, \lambda_{\parallel,\perp,\times}, m$
are even under $A_0\to-A_0,\mu_5\to -\mu_5$, while $  r_\times, \kappa_\times, \gamma_{\parallel,\perp,\times}$
are odd under $A_0\to-A_0,\mu_5\to -\mu_5$. In Appendix~\ref{app:a} we give the full set of constraints for 
all choices of $\Th$.

The equation of motion for $A_{a0}$ again gives $n =0$, which can be used to solve for $A_0$ in terms of $\mu_5, B^2$. Then 
all scalar coefficients, except for $b_{00}$ and $b_{50}$, which are constants,  should be understood as functions of $\mu_5$ and $B^2$. Again terms involving $b_{00}$ and $b_{50}$ should be topological. 

After solving for $A_0$, we find that $n_5$ and $H_i$ satisfy the integrability condition 
\be
  \partial_{\mu_5} H_i = - \partial_{B_i} n_5 \ .
\ee
Equation~\eqref{yeq} gives the constitutive relations for $J^\mu_5$.
From~\eqref{yeq}, we see that $n_5$ is the chiral charge density and  $\ka_{ij}$ is the chiral conductivity tensor.
The term in $J_5^i$ proportional to $c$ gives the Chiral Separation Effect (CSE)~\cite{Son:2004tq,PhysRevD.72.045011} (see also~\cite{Jensen:2013vta,Gorbar:2013upa} for discussion with dynamical fields) in the strong field regime. 
Note that $A_0$ there is in general a nontrivial function of $\mu_5$ and $B^2$, which can be found from the equilibrium free energy density $F (A_0, \mu_5, B^2)$. 
 The $b_{50}$ term is new, as mentioned earlier; it vanishes in a $\sC \sP \sT$ invariant system. The $\lam^-_{ij}$ term  
describes the relation between the chiral current and the dynamical part of the electric field, generalizing the Chiral
Electric Separation Effect (CESE)~\cite{PhysRevLett.110.232302,PhysRevD.91.045001}. $\lam_{ij}^-$ may be called the CESE conductivity tensor~\cite{Kharzeev:2010gd}. 
The last term in~\eqref{yeq} is proportional to the curl of $m \vec B$. It does not contribute to the conservation equation of $J_5^\mu$, but does affect the current itself. It is analogous to the Ampère term for the electric current, but here  with the chiral current. $1/m$ may be thus interpreted as the chiral magnetic permeability, and the term may be referred to as the chiral Ampère effect.

From~\eqref{pos} we find that various transport coefficients satisfy the inequalities
\be\label{pos1}
     r_\parallel >0 , \quad r_\perp >0 \quad  \ka_\parallel \geq \lam_\parallel^2 r_\parallel >0, \quad \kappa_\perp > 0  ,\quad \ka_\perp r_\perp  \geq (\lam_\perp^2+\lam_\times^2) (r_\perp^2 + r_\times^2),
\ee
which hold for all choices of $\Th$. 
For our discussion below it is also convenient to introduce 
\be 
\chi_5  \equiv  {\p n_5 \ov \p \mu_5}, 
\quad 
{\p n_5 \ov \p B_i} = - \p_{\mu_5} H_i = {B_i \ov \mu^2} {\p \mu \ov \p \mu_5}   \equiv \eta_5 B_i 
\ee
where $\chi_5$ can be identified as the chiral susceptibility, and $\eta_5$ the chiral magnetic susceptibility.
Thermodynamic stability of the system also implies that 
\be 
\chi_5 \geq 0, \quad \mu \geq 0, \quad \mu - B \p_B \mu \geq 0,
\ee
while $\eta_5$ can have either sign. 

By design, the equation of motion obtained from varying $\vp_a$ gives~\eqref{an5}, while as in last section, the equations of motion obtained from varying $A_{ai}$ give the constitutive relations of the electric field. More explicitly, they have the form 
\bega \label{ee1}
\chi_5 \p_t \hat \mu_5 + 
\eta_5 B_i  \p_t B_i  -  \partial_i(\kappa_{ij} \partial_j \hat \mu_5) -\partial_i( \lam^-_{ij} \partial_t A_j) =  2 c B_i \p_t A_i , \\
\p_t A_i = - r_{ij} (\ep_{jkl} \p_k H_l+ 2c \hat \mu_5 B_j + \lam^+_{kj} \p_k \hat \mu_5) , \quad \hat \mu_5 \equiv \mu_5 - {b_{00} \ov 2c},
\label{ee2} 
\end{gather}
The term proportional to $c$ in~\eqref{ee2} is the  dynamical field generalization  of the chiral magnetic effect~\cite{PhysRevD.22.3080,Kharzeev:2007jp,Fukushima:2008xe} (see e.g.~\cite{Kharzeev:2013ffa,Kharzeev:2010gd,Landsteiner:2016led}
for reviews). 

After some further manipulations we can write the above equations as 
\bega \label{ee6}
\chi_5 \p_t \hat \mu_5 + (2cB)^2 r_{\parallel}  \hat \mu_5   
+  2 c  r_{\parallel} B_i \ep_{ijk} \p_j  H_k  + \eta_5 B_i \p_t B_i+
2c  B_i ( \partial_i (\hat \mu_5   \Lam^-_{\parallel}) +  \Lam^+_{\parallel}  \p_i  \hat \mu_5)  = \sG, \\
\label{ee7}
 \p_t B_i = - 2c \ep_{ijk} \p_j ( r_{\parallel} \hat \mu_5 B_k) + \sH_{i}  , \qquad \Lam^\pm_{\parallel} = r_\parallel (\lam_\parallel  \pm \ga_\parallel)  \ .
\end{gather}
where $\sG$ and $\sH_i$ denote the second derivative terms whose forms are 
\bega
     \sH_i = -\epsilon_{ijk} \partial_j [r_{kl}(\pp \times \vec H)_l + (r \lam^{+T})_{kl} \partial_l \hat \mu^5),\\
      \sG =  \partial_i(\kappa_{ij} \partial_j \hat \mu_5)  
      - \partial_i[(\lam^- r)_{ij} (\pp\times\vec H)_j + (\lam^- r \lam^{+T})_{ij} \partial_j \hat \mu^5] , 
\end{gather}
with $\lam^{+T}$ denoting the transpose of $\lam^+$. We see that all the leading spatial derivative terms in the above equation come from the anomaly. Note that there can be additional second derivative terms that are proportional to $c$, which are neglected. As we assume $c$ is small, such terms are smaller than the anomaly-independent second derivative terms (which  give rise to diffusion of the chiral charge density and $B_i$) included here.

Now consider homogeneous equilibrium solutions\footnote{In next section we will find more inhomogenous equilibrium solutions.} to~\eqref{ee6}--\eqref{ee7}. We see there are two branches 
\bega \label{equ1}
\hat \mu_5 = {\rm const}, \quad B_i =0 , \\
\hat \mu_5 =0, \quad B_i = {\rm const}  , 
\label{equ2}
\end{gather}
which are connected at $\hat \mu_5 = B_i =0$. The consequence of having $b_{00} \neq 0$ is that in~\eqref{equ2}, there is a nonzero equilibrium value $\mu_5^{\rm eq} = {b_{00} \ov 2c}$.

Note that $\hat \mu_5 \neq 0, B_i \neq 0$ is {\it not} an equilibrium configuration 
due to effects of the anomaly. In fact it is well known that in the presence of a nonzero $B_i$, a nonzero $\hat \mu_5$ will decay (see e.g.~\cite{Das:2022auy} for recent discussions). 
Now suppose we initially have a small nonzero $\hat \mu_5^0$; we then have  
  \be \label{deca}
\hat \mu_5 = \hat \mu_5^0 e^{- \Ga t }, \quad \Ga = {(2c B)^2  r_{\parallel} \ov \chi_5} \biggr|_{\hat \mu^5=0} \ ,
\ee
where both $\chi_5$ and $r_{\parallel}$ are functions of $B$.\footnote{As stressed earlier, $r_{\parallel}$ is not the same as the inverse of conductivity in the presence of a nonzero $B$.} For a finite nonzero $\hat \mu^5_0$, its precise evolution depends on the explicit $\hat \mu_5$-dependence of $\chi_5$ and $r_{\parallel}$, and need not be exponential.

\section{Chiral plasma instability and a nonlinear model}

We now consider small perturbations around the equilibrium configuration~\eqref{equ1}. 
We denote the perturbation of $\hat \mu_5$ as $\nu_5$ and the background value simply as $\hat \mu_5$. 
We can expand all quantities in small $B$; at leading order 
\be\label{uhy}
r_{ij} = r \de_{ij}, \; \ka_{ij} = \ka \de_{ij}, \; \lam_{ij} = \lam \de_{ij}, \;
\ga_{ij} = \ga \de_{ij}, 
\ee
 where $r, \ka, \lam, \ga , \mu, \eta_5$ are all functions of $\hat \mu_5$. 

At linearized level equations~\eqref{ee6}--\eqref{ee7} become decoupled equations 
\bega\label{yh1}
\chi_5 \p_t \nu_5 - \tilde \ka  \nab^2 \nu_5 =  0, \quad \tilde \ka = \ka -  r (\lam^2 - \ga^2) > 0,\\
\p_t   B_i = -  2 c r \hat \mu_5 \ep_{ijk} \p_j  B_k 
+ D_B \nab^2  B_i
 , \quad D_B = {r  \ov \mu},
 \label{yh2} 
\end{gather} 
where~\eqref{yh1} describes diffusion of chiral charge with diffusion constant $D_5 = {\tilde \ka \ov \chi_5}$. 
Equation~\eqref{yh2} can be readily solved by performing a Fourier transform in spatial directions. For sufficiently small $k$, there is an unstable mode growing exponentially with time, which has been termed the chiral plasma instability~\cite{Akamatsu:2013pjd} (see also~\cite{Hirono:2015rla,Kaplan:2016drz,Hattori:2017usa}). 
Taking momentum $\vk = k e_z$ to be along $z$ direction, it can be written~as 
  \be \label{yh3}
 B_x = \sB (t) \cos kz , \quad B_y = \sB (t) \sin kz , \quad \sB (t) = B_0  e^{2cr \hat \mu_5 k t - D_B k^2 t}  \ .
 \ee
The exponential in $\sB(t)$ becomes decaying for $k > 2 c \hat \mu_5 \mu$ and the exponential growth is maximal for 
$k  = c \hat \mu_5 \mu$.\footnote{One may worry that the derivative expansion breaks down when we equate linear and quadratic $k$ terms in the exponential of  $\sB(t)$. This is ok as $\ga \propto c$ is a small parameter. Suppose the coefficients of $k^2$ and $k^3$ terms are controlled by the same scale $\ell$. The cubic and higher terms in $k$ are suppressed provided that 
$ c \hat \mu_5 \mu  \ell \ll 1$.}

To understand the fate of the instability, we need to consider the full nonlinear equations. Here we consider 
a simple model where all transport coefficients are isotropic, as in~\eqref{uhy}, and all the parameters $r, \ka, \lam, \ga , \mu, \eta_5$ are independent of $\mu_5$ and $B$. In this case  equations~\eqref{ee6}--\eqref{ee7}  become 
\bega \label{aa1}
\chi_5 \p_t \hat \mu_5 + (2 c B)^2 r \hat \mu_5+
\eta_5 B_i \p_t B_i  - \tilde \ka  \nab^2 \hat \mu_5    + 2c D_B  \ep_{ikl}  B_i 
 \p_k B_l + 4 c r  \lam B_i \p_i \hat \mu_5  =0
 , \\
\p_t B_i 
= -2 c  r \ep_{ijk} \p_j  (\hat \mu_5 B_k) + D_B \nab^2 B_i  \ .
\label{aa2}
\end{gather}

Now consider a helical magnetic field ansatz of the form 
\be \label{yhl}
\vec B = \sB (t) (\cos kz , \sin kz, 0), \quad \hat \mu_5 = \hat \mu_5 (t), \quad
\nab \times B = -  k \vec B; 
\ee
i.e. $\hat \mu_5$ has no spatial dependence. 
Equations~\eqref{aa1}--\eqref{aa2} then  reduce to 
\bega \label{aa3}
\chi_5 \p_t \hat \mu_5  = -  \le(2c +  \eta_5 k \ri)   (2cr  \hat \mu_5  - D_B k)  \sB^2 
 , \\
\p_t \sB 
=k (2cr  \hat \mu_5  - D_B k) \sB  \ . 
\label{aa4}
\end{gather}
We immediately see that the above equation has a static state 
\be \label{eqv}
\hat \mu_5 = {D_B k \ov 2 cr} = {k \ov 2 c \mu} \equiv \hat \mu_5^{\rm eq} \ ,
\ee
for any value of $\sB$. 

Now let us consider what happens if we have some initial condition 
\be \label{ini}
\hat \mu_5 (t=0) = \hat \mu_5^0, \quad \sB (t=0) = B_0 \ .
\ee
Note that for $ {k \ov 2 c \mu} > \hat \mu_5^0 $, $\sB (t)$ decays with time to zero and linearized analysis is enough. We will thus restrict to those $k$ such that ${k \ov 2 c \mu} < \hat \mu_5^0 $. 
The  qualitative behavior of the evolution can be deduced from~\eqref{aa3}--\eqref{aa4} without solving them explicitly. Suppose $2 c+ \eta_5 k$ is positive, then  the right hand sides of~\eqref{aa3} and~\eqref{aa4} are respectively initially negative and positive. So $\hat\mu_5$ will decrease with time and $\sB (t)$ will increase with time, until $\hat\mu_5$ reaches the equilibrium value~\eqref{eqv} at which point the right hand sides of both equations become zero and the system reaches a steady state. 
The final state will be $\hat\mu_5 =  \hat \mu_5^{\rm eq}$, and $\sB (t)$ is some finite constant. 
If $2 c+ \eta_5 k$ is negative~\footnote{$\eta_5$ can in principle have either sign.}, then for both equations the right hand sides are 
positive, and $\hat \mu_5$ and $\sB$ grow with time indefinitely. The system then appears to have a nonlinear instability. 
Of course, this may result from the fact that our model is too crude. Non-isotropic effects resulting from magnetic fields and dependence of various transport coefficients on $\mu_5, B$ will also be important at some point. Furthermore, when the magnetic field is large enough, temperature fluctuations and velocity effects could also be important.

Equations~\eqref{aa3}--\eqref{aa4} can actually be solved exactly to give 
\bega\label{mu5t}
\hat \mu_5 (t) =\tilde \mu_5^0  { a_k B_0^2 + b k    \ov  b k + a_k B_0^2  e^{(a_k B_0^2 + bk ) t} } +  \hat \mu_5^{\rm eq}, \\ \label{sBt}
\sB (t) =  B_0 {a_k B_0^2 +  bk \ov a_k B_0^2 + bk  e^{ -(a_k B_0^2 + bk) t}} , \\ 
a_k \equiv {2cr \ov \chi_5} \le(2c +  \eta_5 k \ri) , \quad b = 4 c r  \tilde \mu_5^0,  \quad
  \tilde \mu_5^0 \equiv \hat  \mu_5^0 - \hat \mu_5^{\rm eq} \ .
\end{gather}
For $a_k > 0$, the final value of $\sB$ is given by 
\be \label{Binfty}
\sB (\infty) = B_0 + {bk \ov a_k B_0} 
= B_0 + { 2 \chi_5 k  \tilde \mu_5^0 \ov   \le(2c +  \eta_5 k \ri)  B_0 }  \ .
\ee
It is unintuitive that the smaller the initial value $B_0$, the larger $\sB (\infty)$, i.e. $B_0 \to 0$ limit is not smooth.\footnote{This behavior has to do with the $\sB^2$-dependence on the right hand side of~\eqref{aa3}.}  We again expect this feature should not be there when anisotropic effects and dependence on $\mu_5, B$ are included. 

{We note, however, the above discussion deals with a single unstable $k$-mode, while the chiral plasma instability involves a continuum of $k$ values. In particular, if~\eqref{yhl} is perturbed by some modes with $k' \ll k$, the configuration~\eqref{yhl} can be treated as approximately uniform. Then our earlier discussion leading to~\eqref{yh3} can be applied to such perturbations, and there are still instabilities. So it appears likely that~\eqref{equ2} is the only stable configuration and generic perturbations of~\eqref{equ1} will  evolve to~\eqref{equ2}.}

For $\Th=\sC\sP\sT$, we have $b_{00} =0$, meaning that $\hat\mu^5=\mu^5$ and $n_5$---and by extension $\eta_5$---is an odd function of $\mu_5$. So treating $\eta_5$ as a constant is inconsistent. Nevertheless the qualitative discussion below~\eqref{ini} still applies.  

The conclusion we reached here for $2c + \eta_5 k > 0$ is qualitatively similar to an earlier discussion in~\cite{Kaplan:2016drz}, although their nonlinear equations are different. The quantitative conclusions are thus different. For example,~\cite{Kaplan:2016drz} predicts a final value for the magnitude of the helical mode that is independent of initial conditions.

\section{\Dcmw\ in the strong field regime}  \label{sec:dcmw}

Now let us consider small perturbations around the branch~\eqref{equ2}. 
We take the equilibrium magnetic field to be $\vec B = B_0 e_z$, and denote its perturbation as $\vec b$, i.e. $\vec B = B_0 e_z + \vec b$. To linear order in $\hat \mu_5$ and $\vec b$, equations for~\eqref{ee6} and~\eqref{ee7} become 
\bega  \label{ss1}
\chi_5 \p_t \hat \mu_5 + (2cB_0)^2 r_{\parallel}  \hat \mu_5   +  4 c B_0 r_{\parallel} \lam_{\parallel} \p_z \hat \mu_5 
+  2 c  r_{\parallel} H_0 \ep_{ab} \p_a  b_b - \sD_{ij\al} \p_i \p_j \hat b_\al =0, \\
  \label{ss2}
\p_t b_i = - 2c  B_0 r_{\parallel}\de_i^a \ep_{ab} \p_b   \hat \mu_5 - \sE_{ijk\al} \p_j \p_k \hat b_\al ,
\quad H_0 = - {\p F \ov \p B_0} , \quad \hat b_\al= (\hat \mu_5, b_i) , \quad a, b = x, y  \ .
\end{gather}
The explicit expressions of $\sD_{ij\al}$ and $\sE_{ijk \al}$ are a bit long, which we will not need below. 
They are independent of anomaly coefficient $c$ and describe coupled diffusions of chiral charge density and $b_i$. 
Going to Fourier space for both $t$ and $\vx$ we find the equation of $\hat \mu_5$ gives the dispersion relation 
\be\label{jen}
\om = -i \Ga + v_z k_z + O(k^2_{z,\perp})   , \quad
\Ga \equiv {(2 c B_0)^2 r_\parallel \ov \chi_5}, \quad v_z \equiv { 4c B_0 r_\parallel  \lam_\parallel  \ov \chi_5}.
\ee
Denote the initial value of $\hat \mu_5$ (i.e. at $t=0$) by $f_0 (\vx)$; 
we then find 
\bega 
\hat \mu_5 (\vx, t)  
= f_0 (\vec x_\perp, z - v_z t) e^{-\Ga t}  \ .
 \end{gather}
We thus find, in addition to an overall exponential decay, along the $z$-direction there is a propagating wave with velocity $v_z$. 
As $r_\parallel$ is non-negative, the sign of $v_z$ is determined from that of $\lam_\parallel$. 
There is a propagating wave only along the positive $z$ or negative $z$ direction depending on the sign of $v_z$; i.e. the propagation is chiral. {Including $O(k^2)$ terms, there are also diffusion in $x, y$ directions, and $k$-dependent attenuation of the wave.}

Now consider~\eqref{ss2}. At leading order in spatial derivatives, the contribution to $\p_t b_z$ comes from $O(k^2)$ terms, and the dominant behavior should be diffusion.  For $b_a$ it is determined by the spatial derivative of $\hat \mu_5$; we thus find that 
\be 
\p_t b_a =  - 2cB_0  r_\parallel  \ep_{ab} \p_b  f_0 (\vec x_\perp, z - v_z t) e^{-\Ga t} ,
\ee
which gives 
\be 
b_a (t, \vx) = b_a^0 (\vx) - 2cB_0  r_\parallel  \ep_{ab} \p_b  \le(g_0 (\vec x_\perp, z - v_z t) e^{-\Ga t}  - g_0 (\vx) \ri) ,
\ee
where $g_0 (\vx)$ is a function which can be obtained from $f_0$ by an integral transform. 
Thus $b_a (t, \vx)$ also exhibit a decaying chiral wave. 
When the second derivative terms are included, there are again diffusions under which the initial perturbation $b_a^0 (\vx)$ will evolve to zero. 

{The chiral wave here is reminiscent of the chiral magnetic wave (CMW) discussed in~\cite{Kharzeev:2010gd} with non-dynamical external fields (which we briefly review in Appendix~\ref{app:cmw}). But it is fundamentally a different effect.\footnote{{
We thank Ho-Ung Yee for a discussion on this.} An earlier discussion of chiral MHD~\cite{Hattori:2017usa}  appears not to find CMW when magnetic fields are dynamical.} More explicitly, when the $U(1)$ symmetry is global,  fluctuations in the electric charge density induce chiral currents via the CSE and fluctuations in the chiral charge density induce electric currents via the CME, leading to the CMW.
 By contrast, when the $U(1)$ symmetry is gauged, as discussed earlier, the electric charge density 
is always zero, $n=0$.} As a result, the CSE no longer permits electric charge density fluctuations to induce chiral currents. 
{Instead, now CESE enables electric field fluctuations to induce chiral current fluctuations, and 
the electric field fluctuations are in turn sourced by fluctuations in the chiral charge density (chiral chemical potential) through 
the dynamical version of the CME, which leads to the chiral wave in the chiral charge density. 
While CMW in~\cite{Kharzeev:2010gd} has two chiral waves traveling in opposite directions, there is only one chiral wave in the chiral charge density (which also sources a chiral wave in the magnetic field).}
 We will refer to this new chiral wave as the \dcmw. 
 
 
{We stress that the velocity~\eqref{jen} is proportional to $\lam_\para$, which vanishes when the system conserves $\sC$. For a parity invariant system, $\lam_\para$ should be odd in $\mu_5$.} 
 


\section{Discussions}

We have given a general formulation of EFTs for  electromagnetic fields in general media.
The specific form of the action depends on the choice of a derivative expansion scheme. 
 Here we focused on the regime for MHD, and restrict to the case where flow velocities and temperature 
 variations can be neglected. We then considered the situation with chiral charged matter and ABJ anomaly. 
 The resulting theory gives dynamical field generalization of the chiral magnetic effect, the chiral separation effect, and the chiral electric separation effect. Linearizing around the equilibrium configuration of a constant magnetic field reveals that the chiral magnetic wave does not survive {dynamical electromagnetic fields, but a new \dcmw\ emerges.  We predict its wave velocity.} We also present a simple full nonlinear model for the chiral plasma instability. 
 
Understanding possible field-dependence of various transport parameters in realistic applications  (e.g. heavy ion collisions,  Dirac and Weyl semimetals, electroweak plasmas and neutron stars) is clearly of crucial importance. 
 It would be instructive  to find toy models in which such field-dependences 
can be explicitly worked out, and study the physical implications of such dependence
on the evolution of the MHD equations. Holographic systems should provide a good such laboratory.

\begin{acknowledgments}
We would like to thank Paolo Glorioso, Ho-Ung Yee, Yin Yi for discussions and Nabil Iqbal for correspondence. This work is supported by the Office of High Energy Physics of U.S. Department of Energy under grant Contract Number  DE-SC0012567 and the ALFA foundation. 
\end{acknowledgments}

\appendix

\section{Choices of $\Theta$ for dynamical KMS symmetry}\label{app:a}

Consider the most general Lagrangian for chiral anomalous plasma up to weight-two in the derivative expansion such that under a dynamical KMS transformation (irrespective of choice of $\Theta$) all terms that do not involve $a$-type fields are total derivatives,\footnote{For simplicity, we work to linear order in $\sB_{ai}$ and include the minimal factors of $\sB_i$ necessary to ensure dynamical KMS symmetry is satisfied. Higher-order terms in $\sB_i$ may be important for correlation functions, but they alter neither the equations of motion nor the constitutive relation for $J^\mu_5$ and thus fall outside the scope of this work. } 
\be\begin{split}\label{inv T}
     \sL  = \hat \sL_\text{anom} + b_{00} B_i A_{ai} + b_{50} (\sB_i A_{ai} + B_i C_{ai}) + n A_{a0} + n^5 C_{a0} - H^i B_{ai} \\
     +i (r^{-1})_{ij} A_{ai}(T_0 A_{aj} + i \p_t A_j ) + i \kappa_{ij} \partial_i \varphi_a (T_0 \partial_j \varphi_a +i\partial_j \mu_5) \\
   +i  \lam_{ij}  (  (T_0 \partial_i \varphi_a +i\partial_i \mu_5) A_{aj}+ \p_i \vp_a  (T_0 A_{aj} + i \p_t A_j )) \\
+ \gamma_{ij} ( \partial_i \varphi_a \partial_t A_j -  \partial_i \mu_5 A_{aj}) + m (B_i \sB_{ai} + B_{ai} \sB_i) + m_a B_i \sB_i \ ,
\end{split} \ee
for which $b_{00}, b_{05}$ are constant,~\eqref{desc} is satisfied and 
\bega
     m_a = {\p m \ov  \p A_0}A_{a0} + {\p m \ov \p C_0} C_{a0} + {\p m \ov \p B_i} B_{ai},\\
     \hat \sL_\text{anom} =  -2 c  (- \p_t \vec A\cdot \vec B\varphi_a +  (\dot \varphi \vec B - \p_t \vec A \times\vec \sA)
\cdot\vec A_a + A_0 (\vec \sA \cdot \vec B_a +\vec B\cdot \vec \sA_a)
+  \vec B\cdot \vec \sA A_{a0} ) \ .
\end{gather}
We have defined $\hat \sL_\text{anom}$ to be the minimal set of terms containing the anomaly that is invariant under all symmetries; it has no free parameters and is thus independent of our choice of $\Th$. 

The form of this leading-order Lagrangian is necessary, but not sufficient to ensure that dynamical KMS symmetry is imposed. We will now investigate how the various choices of $\Th$ affect the action; for each choice, we must impose certain additional constraints as follows.
\ben
     \item $\Th=\sT$: The only constraints are $\lambda_\times =\gamma_{\parallel,\perp}=0$. 
     \item $\Th=\sP\sT$: $ b_{00}=0 $ and the coefficients 
     \be
          F,\quad r_{\parallel,\perp,\times},\quad \kappa_{\parallel,\perp,\times},\quad \lambda_\times \quad \gamma_{\parallel,\perp},\quad m
     \ee
     are all even functions of $\mu^5$ while $\lambda_{\parallel,\perp}$ and $\gamma_\times$ are all odd functions. 
     \item $\Th=\sC\sT$: $b_{50}=0$ and that the coefficients 
     \be
          F, \quad r_{\parallel,\perp},\quad \kappa_{\parallel,\perp}, \quad \gamma_{\parallel,\perp,\times}
     \ee
     are all even functions of $A_0$ while
     \be\label{CT coef}
          r_\times,\quad \kappa_\times,\quad \lambda_{\parallel,\perp,\times},\quad m
     \ee
     are all odd functions. Further, notice that the equation of of motion for $A_0$ is $n=-\partial_{A_0}F=0$. As $F$ is an even function of $A_0$ this equation is solved by $A_0=0$, meaning that on the the equations of motion,~\eqref{CT coef} all vanish.
     \item $\Th=\sC\sP\sT$: Both topological coefficients vanish, $b_{00}=b_{50}=0$. And the coefficients 
     \be\label{CPT coef1}
          F, \quad r_{\parallel,\perp},\quad \kappa_{\parallel,\perp},\quad \lambda_{\parallel,\perp,\times},\quad m
     \ee
     are even under $A_0\to-A_0,\mu^5\to -\mu^5$, while
     \be\label{CPT coef2}
          r_\times,\quad \kappa_\times,\quad \gamma_{\parallel,\perp,\times}
     \ee
     are all odd. The equation of motion for $A_{a0}$ is $n=-\partial_{A_0}F=0$, which can be solved for $A_0$ in terms of $B$ and $\mu^5$. Because $F$ is even in $A_0,\mu^5$, the expression for $A_0$ must be odd in $\mu^5$. As a result, on shell, \eqref{CPT coef1} are understood to be even functions of $\mu^5$ while \eqref{CPT coef2} are odd. 
\een

Interestingly, unlike the case of ordinary plasmas, the various choices of $\Th$ do not correspond to more symmetric versions of the $\Th=\sT$ case.

\section{Comparison with the results of~\cite{Vardhan:2022wxz}} \label{app:b}

Here we compare~\eqref{teg} with the corresponding expressions in~\cite{Vardhan:2022wxz}. 
From~\eqref{teg} we have 
\bea
E_i  &=& \nab A_0 - \p_0 A_i  =  \p_i f  + (r_\para h_i h_j + r_\perp \De_{ij} + r_\times \ep_{ijk} h_k ) ((\nab \times \vec H)_j - b_0 B_j) \\
& = & \p_i f   - r_\para b_0 B_i + (r_\para h_i h_j + r_\perp \De_{ij} + r_\times \ep_{ijk} h_k ) (\nab \times \vec H)_j 
 \eea
where $H_i = {B_i \ov \mu}$. Note that 
\be 
\nab \times \vec H = \nab \times \le({\vec B \ov \mu} \ri) = {\vec j_B \ov \mu} , \quad 
{\vec j}_{B} \equiv {\vec j}  - \pp \ln \mu \times {\vec B}  , \quad \vec j \equiv \nab \times \vec B \ .
\ee
We then find that 
\bega \label{ywe}
{\vec E} =\pp f - b_0 r_\parallel  {\vec B}   + c_{\eta}\,  {\vec j}_B +c_a \, ({\vec B} \cdot {\vec j}) \, {\vec B} + c_H \, {\vec j}_B\times {\vec B}  , \\ 
 c_\eta = {r_\perp\ov \mu} ,\quad c_a =  \frac{r_\parallel-r_\perp}{\mu B^2}  \quad c_H = \frac{ r_\times}{\mu B} \ .
\end{gather} 
Equation~\eqref{ywe} can also be written as 
 \begin{align}
{\vec E}  & =\pp f - b_0 r_\parallel  {\vec B}  +
 \eta \, {\vec j} + c_a \, ({\vec j} \times {\vec B})\times{ \vec B} + c_H \, {\vec j} \times {\vec B} 
 \cr
 &- c_H (\pp \ln a \times {\vec B}) \times {\vec B} - c_{\eta} \pp  \ln a \times {\vec B}  ,
\qquad \eta = { r_\parallel \ov \mu} \ .
\label{ouC}
 \end{align} 
Equations~\eqref{ywe} and~\eqref{ouC} are exactly the same as equations (6) and (9) of~\cite{Vardhan:2022wxz} (except that $b_0$-term is missing there) with the identifications with parameters there 
\be
\mu (B^2)= a (G^2), \quad r_\perp = 2 \beta_0 d, \quad r_\times = -{p B \ov \mu} , \quad r_\para = 
2 \beta_0 d+  \frac{4 \beta_0 \tilde d B^2}{\mu^2}  \ .
\ee
For background $\vec B = B_0 e_z$, the transport coefficients computed in~\cite{Vardhan:2022wxz} can be expressed in our notation by
\be
     \chi_\parallel = {\mu^2 \ov \mu - B \p_B \mu} ,\quad \chi_\perp = \mu,\quad \sigma^\perp_1 = r_\perp,\quad \sigma_2^\perp = {r_\times} ,\quad \sigma^\parallel = r_\parallel \ . 
\ee

 The fact that the partial Higgs approach and the 1-form symmetry method yield the same equations of motion is no accident; there is a duality transformation that can convert one action into the other~\cite{LandryLiu}. 
 This also explains the missing $b_0$ term there, as this term is non-local in the one-form symmetry formulation.


\section{Chiral magnetic wave (CMW)}  \label{app:cmw} 

To contrast with the discussion of the \dcmw\ of Sec.~\ref{sec:dcmw}, 
here we give a brief review of the CMW  discussed in~\cite{Kharzeev:2010gd} with non-dynamical external fields. 
Suppose the system has a constant external magnetic field $\vec B = B e_z$ in $z$-direction.  
From the (non-dynamical) CSE and CME effects, the vector and chiral currents are respectively given by
\be 
\vec J = - 2 c \mu_5 \vec B, \quad \vec J_5 = -2 c \mu \vec B ,
\ee
where $\mu$ and $\mu_5$ are the chemical potentials for the charge and chiral charge densities.\footnote{{Note that $\mu$ here is in no way related to the magnetic permability.}} 
Now consider small perturbations of $\mu$ and $\mu_5$, which induce perturbations of the corresponding charge densities 
\be 
J^0 = \chi \de\mu , \quad J^0_5 = \chi_5 \de \mu_5 \ .
\ee
In the absence of an external electric field, the conservation equations of $J^\mu$ and $J^\mu_5$ lead to 
\be 
\chi \p_t \mu - 2 cB \p_z \mu_5  = 0 , \quad \chi_5 \p_t \mu_5 - 2 c B \p_z \mu  = 0, 
\ee
which give rise to {\it two} chiral waves traveling in opposite directions, with dispersion relations 
\be 
\om = \pm {2cB \ov \sqrt{\chi \chi_5}} k_z  \ .
\ee

\bibliographystyle{apsrev4-1}
\bibliography{biblio}{}

\begin{thebibliography}{53}%
\makeatletter
\providecommand \@ifxundefined [1]{%
 \@ifx{#1\undefined}
}%
\providecommand \@ifnum [1]{%
 \ifnum #1\expandafter \@firstoftwo
 \else \expandafter \@secondoftwo
 \fi
}%
\providecommand \@ifx [1]{%
 \ifx #1\expandafter \@firstoftwo
 \else \expandafter \@secondoftwo
 \fi
}%
\providecommand \natexlab [1]{#1}%
\providecommand \enquote  [1]{``#1''}%
\providecommand \bibnamefont  [1]{#1}%
\providecommand \bibfnamefont [1]{#1}%
\providecommand \citenamefont [1]{#1}%
\providecommand \href@noop [0]{\@secondoftwo}%
\providecommand \href [0]{\begingroup \@sanitize@url \@href}%
\providecommand \@href[1]{\@@startlink{#1}\@@href}%
\providecommand \@@href[1]{\endgroup#1\@@endlink}%
\providecommand \@sanitize@url [0]{\catcode `\\12\catcode `\$12\catcode
  `\&12\catcode `\#12\catcode `\^12\catcode `\_12\catcode `\%12\relax}%
\providecommand \@@startlink[1]{}%
\providecommand \@@endlink[0]{}%
\providecommand \url  [0]{\begingroup\@sanitize@url \@url }%
\providecommand \@url [1]{\endgroup\@href {#1}{\urlprefix }}%
\providecommand \urlprefix  [0]{URL }%
\providecommand \Eprint [0]{\href }%
\providecommand \doibase [0]{http://dx.doi.org/}%
\providecommand \selectlanguage [0]{\@gobble}%
\providecommand \bibinfo  [0]{\@secondoftwo}%
\providecommand \bibfield  [0]{\@secondoftwo}%
\providecommand \translation [1]{[#1]}%
\providecommand \BibitemOpen [0]{}%
\providecommand \bibitemStop [0]{}%
\providecommand \bibitemNoStop [0]{.\EOS\space}%
\providecommand \EOS [0]{\spacefactor3000\relax}%
\providecommand \BibitemShut  [1]{\csname bibitem#1\endcsname}%
\let\auto@bib@innerbib\@empty
\bibitem [{\citenamefont {Vardhan}\ \emph {et~al.}(2022)\citenamefont
  {Vardhan}, \citenamefont {Grozdanov}, \citenamefont {Leutheusser},\ and\
  \citenamefont {Liu}}]{Vardhan:2022wxz}%
  \BibitemOpen
  \bibfield  {author} {\bibinfo {author} {\bibfnamefont {S.}~\bibnamefont
  {Vardhan}}, \bibinfo {author} {\bibfnamefont {S.}~\bibnamefont {Grozdanov}},
  \bibinfo {author} {\bibfnamefont {S.}~\bibnamefont {Leutheusser}}, \ and\
  \bibinfo {author} {\bibfnamefont {H.}~\bibnamefont {Liu}},\ }\href@noop {} {\
   (\bibinfo {year} {2022})},\ \Eprint {http://arxiv.org/abs/2207.01636}
  {arXiv:2207.01636 [astro-ph.HE]} \BibitemShut {NoStop}%
\bibitem [{\citenamefont {Gaiotto}\ \emph {et~al.}(2015)\citenamefont
  {Gaiotto}, \citenamefont {Kapustin}, \citenamefont {Seiberg},\ and\
  \citenamefont {Willett}}]{Gaiotto:2014kfa}%
  \BibitemOpen
  \bibfield  {author} {\bibinfo {author} {\bibfnamefont {D.}~\bibnamefont
  {Gaiotto}}, \bibinfo {author} {\bibfnamefont {A.}~\bibnamefont {Kapustin}},
  \bibinfo {author} {\bibfnamefont {N.}~\bibnamefont {Seiberg}}, \ and\
  \bibinfo {author} {\bibfnamefont {B.}~\bibnamefont {Willett}},\ }\href
  {\doibase 10.1007/JHEP02(2015)172} {\bibfield  {journal} {\bibinfo  {journal}
  {JHEP}\ }\textbf {\bibinfo {volume} {02}},\ \bibinfo {pages} {172} (\bibinfo
  {year} {2015})},\ \Eprint {http://arxiv.org/abs/1412.5148} {arXiv:1412.5148
  [hep-th]} \BibitemShut {NoStop}%
\bibitem [{\citenamefont {Grozdanov}\ \emph {et~al.}(2017)\citenamefont
  {Grozdanov}, \citenamefont {Hofman},\ and\ \citenamefont
  {Iqbal}}]{Grozdanov:2016tdf}%
  \BibitemOpen
  \bibfield  {author} {\bibinfo {author} {\bibfnamefont {S.}~\bibnamefont
  {Grozdanov}}, \bibinfo {author} {\bibfnamefont {D.~M.}\ \bibnamefont
  {Hofman}}, \ and\ \bibinfo {author} {\bibfnamefont {N.}~\bibnamefont
  {Iqbal}},\ }\href {\doibase 10.1103/PhysRevD.95.096003} {\bibfield  {journal}
  {\bibinfo  {journal} {Phys. Rev.}\ }\textbf {\bibinfo {volume} {D95}},\
  \bibinfo {pages} {096003} (\bibinfo {year} {2017})},\ \Eprint
  {http://arxiv.org/abs/1610.07392} {arXiv:1610.07392 [hep-th]} \BibitemShut
  {NoStop}%
\bibitem [{\citenamefont {Landsteiner}(2016)}]{Landsteiner:2016led}%
  \BibitemOpen
  \bibfield  {author} {\bibinfo {author} {\bibfnamefont {K.}~\bibnamefont
  {Landsteiner}},\ }\href {\doibase 10.5506/APhysPolB.47.2617} {\bibfield
  {journal} {\bibinfo  {journal} {Acta Phys. Polon. B}\ }\textbf {\bibinfo
  {volume} {47}},\ \bibinfo {pages} {2617} (\bibinfo {year} {2016})},\ \Eprint
  {http://arxiv.org/abs/1610.04413} {arXiv:1610.04413 [hep-th]} \BibitemShut
  {NoStop}%
\bibitem [{\citenamefont {Kharzeev}\ \emph {et~al.}(2016)\citenamefont
  {Kharzeev}, \citenamefont {Liao}, \citenamefont {Voloshin},\ and\
  \citenamefont {Wang}}]{Kharzeev:2015znc}%
  \BibitemOpen
  \bibfield  {author} {\bibinfo {author} {\bibfnamefont {D.~E.}\ \bibnamefont
  {Kharzeev}}, \bibinfo {author} {\bibfnamefont {J.}~\bibnamefont {Liao}},
  \bibinfo {author} {\bibfnamefont {S.~A.}\ \bibnamefont {Voloshin}}, \ and\
  \bibinfo {author} {\bibfnamefont {G.}~\bibnamefont {Wang}},\ }\href {\doibase
  10.1016/j.ppnp.2016.01.001} {\bibfield  {journal} {\bibinfo  {journal} {Prog.
  Part. Nucl. Phys.}\ }\textbf {\bibinfo {volume} {88}},\ \bibinfo {pages} {1}
  (\bibinfo {year} {2016})},\ \Eprint {http://arxiv.org/abs/1511.04050}
  {arXiv:1511.04050 [hep-ph]} \BibitemShut {NoStop}%
\bibitem [{\citenamefont {Golkar}\ and\ \citenamefont
  {Son}(2015)}]{Golkar:2012kb}%
  \BibitemOpen
  \bibfield  {author} {\bibinfo {author} {\bibfnamefont {S.}~\bibnamefont
  {Golkar}}\ and\ \bibinfo {author} {\bibfnamefont {D.~T.}\ \bibnamefont
  {Son}},\ }\href {\doibase 10.1007/JHEP02(2015)169} {\bibfield  {journal}
  {\bibinfo  {journal} {JHEP}\ }\textbf {\bibinfo {volume} {02}},\ \bibinfo
  {pages} {169} (\bibinfo {year} {2015})},\ \Eprint
  {http://arxiv.org/abs/1207.5806} {arXiv:1207.5806 [hep-th]} \BibitemShut
  {NoStop}%
\bibitem [{\citenamefont {Hou}\ \emph {et~al.}(2012)\citenamefont {Hou},
  \citenamefont {Liu},\ and\ \citenamefont {Ren}}]{Hou:2012xg}%
  \BibitemOpen
  \bibfield  {author} {\bibinfo {author} {\bibfnamefont {D.-F.}\ \bibnamefont
  {Hou}}, \bibinfo {author} {\bibfnamefont {H.}~\bibnamefont {Liu}}, \ and\
  \bibinfo {author} {\bibfnamefont {H.-c.}\ \bibnamefont {Ren}},\ }\href
  {\doibase 10.1103/PhysRevD.86.121703} {\bibfield  {journal} {\bibinfo
  {journal} {Phys. Rev. D}\ }\textbf {\bibinfo {volume} {86}},\ \bibinfo
  {pages} {121703} (\bibinfo {year} {2012})},\ \Eprint
  {http://arxiv.org/abs/1210.0969} {arXiv:1210.0969 [hep-th]} \BibitemShut
  {NoStop}%
\bibitem [{\citenamefont {Jensen}\ \emph {et~al.}(2013)\citenamefont {Jensen},
  \citenamefont {Kovtun},\ and\ \citenamefont {Ritz}}]{Jensen:2013vta}%
  \BibitemOpen
  \bibfield  {author} {\bibinfo {author} {\bibfnamefont {K.}~\bibnamefont
  {Jensen}}, \bibinfo {author} {\bibfnamefont {P.}~\bibnamefont {Kovtun}}, \
  and\ \bibinfo {author} {\bibfnamefont {A.}~\bibnamefont {Ritz}},\ }\href
  {\doibase 10.1007/JHEP10(2013)186} {\bibfield  {journal} {\bibinfo  {journal}
  {JHEP}\ }\textbf {\bibinfo {volume} {10}},\ \bibinfo {pages} {186} (\bibinfo
  {year} {2013})},\ \Eprint {http://arxiv.org/abs/1307.3234} {arXiv:1307.3234
  [hep-th]} \BibitemShut {NoStop}%
\bibitem [{\citenamefont {Gorbar}\ \emph {et~al.}(2013)\citenamefont {Gorbar},
  \citenamefont {Miransky}, \citenamefont {Shovkovy},\ and\ \citenamefont
  {Wang}}]{Gorbar:2013upa}%
  \BibitemOpen
  \bibfield  {author} {\bibinfo {author} {\bibfnamefont {E.~V.}\ \bibnamefont
  {Gorbar}}, \bibinfo {author} {\bibfnamefont {V.~A.}\ \bibnamefont
  {Miransky}}, \bibinfo {author} {\bibfnamefont {I.~A.}\ \bibnamefont
  {Shovkovy}}, \ and\ \bibinfo {author} {\bibfnamefont {X.}~\bibnamefont
  {Wang}},\ }\href {\doibase 10.1103/PhysRevD.88.025025} {\bibfield  {journal}
  {\bibinfo  {journal} {Phys. Rev. D}\ }\textbf {\bibinfo {volume} {88}},\
  \bibinfo {pages} {025025} (\bibinfo {year} {2013})},\ \Eprint
  {http://arxiv.org/abs/1304.4606} {arXiv:1304.4606 [hep-ph]} \BibitemShut
  {NoStop}%
\bibitem [{\citenamefont {Hirono}\ \emph {et~al.}(2015)\citenamefont {Hirono},
  \citenamefont {Kharzeev},\ and\ \citenamefont {Yin}}]{Hirono:2015rla}%
  \BibitemOpen
  \bibfield  {author} {\bibinfo {author} {\bibfnamefont {Y.}~\bibnamefont
  {Hirono}}, \bibinfo {author} {\bibfnamefont {D.}~\bibnamefont {Kharzeev}}, \
  and\ \bibinfo {author} {\bibfnamefont {Y.}~\bibnamefont {Yin}},\ }\href
  {\doibase 10.1103/PhysRevD.92.125031} {\bibfield  {journal} {\bibinfo
  {journal} {Phys. Rev. D}\ }\textbf {\bibinfo {volume} {92}},\ \bibinfo
  {pages} {125031} (\bibinfo {year} {2015})},\ \Eprint
  {http://arxiv.org/abs/1509.07790} {arXiv:1509.07790 [hep-th]} \BibitemShut
  {NoStop}%
\bibitem [{\citenamefont {Figueroa}\ and\ \citenamefont
  {Shaposhnikov}(2018)}]{Figueroa:2017hun}%
  \BibitemOpen
  \bibfield  {author} {\bibinfo {author} {\bibfnamefont {D.~G.}\ \bibnamefont
  {Figueroa}}\ and\ \bibinfo {author} {\bibfnamefont {M.}~\bibnamefont
  {Shaposhnikov}},\ }\href {\doibase 10.1007/JHEP04(2018)026} {\bibfield
  {journal} {\bibinfo  {journal} {JHEP}\ }\textbf {\bibinfo {volume} {04}},\
  \bibinfo {pages} {026} (\bibinfo {year} {2018})},\ \bibinfo {note} {[Erratum:
  JHEP 07, 217 (2020)]},\ \Eprint {http://arxiv.org/abs/1707.09967}
  {arXiv:1707.09967 [hep-ph]} \BibitemShut {NoStop}%
\bibitem [{\citenamefont {Figueroa}\ \emph {et~al.}(2019)\citenamefont
  {Figueroa}, \citenamefont {Florio},\ and\ \citenamefont
  {Shaposhnikov}}]{Figueroa:2019jsi}%
  \BibitemOpen
  \bibfield  {author} {\bibinfo {author} {\bibfnamefont {D.~G.}\ \bibnamefont
  {Figueroa}}, \bibinfo {author} {\bibfnamefont {A.}~\bibnamefont {Florio}}, \
  and\ \bibinfo {author} {\bibfnamefont {M.}~\bibnamefont {Shaposhnikov}},\
  }\href {\doibase 10.1007/JHEP10(2019)142} {\bibfield  {journal} {\bibinfo
  {journal} {JHEP}\ }\textbf {\bibinfo {volume} {10}},\ \bibinfo {pages} {142}
  (\bibinfo {year} {2019})},\ \Eprint {http://arxiv.org/abs/1904.11892}
  {arXiv:1904.11892 [hep-th]} \BibitemShut {NoStop}%
\bibitem [{\citenamefont {Das}\ \emph {et~al.}(2022)\citenamefont {Das},
  \citenamefont {Gregory},\ and\ \citenamefont {Iqbal}}]{Das:2022auy}%
  \BibitemOpen
  \bibfield  {author} {\bibinfo {author} {\bibfnamefont {A.}~\bibnamefont
  {Das}}, \bibinfo {author} {\bibfnamefont {R.}~\bibnamefont {Gregory}}, \ and\
  \bibinfo {author} {\bibfnamefont {N.}~\bibnamefont {Iqbal}},\ }\href@noop {}
  {\  (\bibinfo {year} {2022})},\ \Eprint {http://arxiv.org/abs/2205.03619}
  {arXiv:2205.03619 [hep-th]} \BibitemShut {NoStop}%
\bibitem [{\citenamefont {Nair}\ \emph {et~al.}(2012)\citenamefont {Nair},
  \citenamefont {Ray},\ and\ \citenamefont {Roy}}]{Nair:2011mk}%
  \BibitemOpen
  \bibfield  {author} {\bibinfo {author} {\bibfnamefont {V.~P.}\ \bibnamefont
  {Nair}}, \bibinfo {author} {\bibfnamefont {R.}~\bibnamefont {Ray}}, \ and\
  \bibinfo {author} {\bibfnamefont {S.}~\bibnamefont {Roy}},\ }\href {\doibase
  10.1103/PhysRevD.86.025012} {\bibfield  {journal} {\bibinfo  {journal} {Phys.
  Rev. D}\ }\textbf {\bibinfo {volume} {86}},\ \bibinfo {pages} {025012}
  (\bibinfo {year} {2012})},\ \Eprint {http://arxiv.org/abs/1112.4022}
  {arXiv:1112.4022 [hep-th]} \BibitemShut {NoStop}%
\bibitem [{\citenamefont {Capasso}\ \emph {et~al.}(2013)\citenamefont
  {Capasso}, \citenamefont {Nair},\ and\ \citenamefont
  {Tekel}}]{Capasso:2013jva}%
  \BibitemOpen
  \bibfield  {author} {\bibinfo {author} {\bibfnamefont {D.}~\bibnamefont
  {Capasso}}, \bibinfo {author} {\bibfnamefont {V.~P.}\ \bibnamefont {Nair}}, \
  and\ \bibinfo {author} {\bibfnamefont {J.}~\bibnamefont {Tekel}},\ }\href
  {\doibase 10.1103/PhysRevD.88.085025} {\bibfield  {journal} {\bibinfo
  {journal} {Phys. Rev. D}\ }\textbf {\bibinfo {volume} {88}},\ \bibinfo
  {pages} {085025} (\bibinfo {year} {2013})},\ \Eprint
  {http://arxiv.org/abs/1307.7610} {arXiv:1307.7610 [hep-th]} \BibitemShut
  {NoStop}%
\bibitem [{\citenamefont {Monteiro}\ \emph {et~al.}(2015)\citenamefont
  {Monteiro}, \citenamefont {Abanov},\ and\ \citenamefont
  {Nair}}]{Monteiro:2014wsa}%
  \BibitemOpen
  \bibfield  {author} {\bibinfo {author} {\bibfnamefont {G.~M.}\ \bibnamefont
  {Monteiro}}, \bibinfo {author} {\bibfnamefont {A.~G.}\ \bibnamefont
  {Abanov}}, \ and\ \bibinfo {author} {\bibfnamefont {V.~P.}\ \bibnamefont
  {Nair}},\ }\href {\doibase 10.1103/PhysRevD.91.125033} {\bibfield  {journal}
  {\bibinfo  {journal} {Phys. Rev. D}\ }\textbf {\bibinfo {volume} {91}},\
  \bibinfo {pages} {125033} (\bibinfo {year} {2015})},\ \Eprint
  {http://arxiv.org/abs/1410.4833} {arXiv:1410.4833 [hep-th]} \BibitemShut
  {NoStop}%
\bibitem [{\citenamefont {Boyarsky}\ \emph {et~al.}(2015)\citenamefont
  {Boyarsky}, \citenamefont {Frohlich},\ and\ \citenamefont
  {Ruchayskiy}}]{Boyarsky:2015faa}%
  \BibitemOpen
  \bibfield  {author} {\bibinfo {author} {\bibfnamefont {A.}~\bibnamefont
  {Boyarsky}}, \bibinfo {author} {\bibfnamefont {J.}~\bibnamefont {Frohlich}},
  \ and\ \bibinfo {author} {\bibfnamefont {O.}~\bibnamefont {Ruchayskiy}},\
  }\href {\doibase 10.1103/PhysRevD.92.043004} {\bibfield  {journal} {\bibinfo
  {journal} {Phys. Rev. D}\ }\textbf {\bibinfo {volume} {92}},\ \bibinfo
  {pages} {043004} (\bibinfo {year} {2015})},\ \Eprint
  {http://arxiv.org/abs/1504.04854} {arXiv:1504.04854 [hep-ph]} \BibitemShut
  {NoStop}%
\bibitem [{\citenamefont {Pavlovi\'c}\ \emph {et~al.}(2017)\citenamefont
  {Pavlovi\'c}, \citenamefont {Leite},\ and\ \citenamefont
  {Sigl}}]{Pavlovic:2016gac}%
  \BibitemOpen
  \bibfield  {author} {\bibinfo {author} {\bibfnamefont {P.}~\bibnamefont
  {Pavlovi\'c}}, \bibinfo {author} {\bibfnamefont {N.}~\bibnamefont {Leite}}, \
  and\ \bibinfo {author} {\bibfnamefont {G.}~\bibnamefont {Sigl}},\ }\href
  {\doibase 10.1103/PhysRevD.96.023504} {\bibfield  {journal} {\bibinfo
  {journal} {Phys. Rev. D}\ }\textbf {\bibinfo {volume} {96}},\ \bibinfo
  {pages} {023504} (\bibinfo {year} {2017})},\ \Eprint
  {http://arxiv.org/abs/1612.07382} {arXiv:1612.07382 [astro-ph.CO]}
  \BibitemShut {NoStop}%
\bibitem [{\citenamefont {Giovannini}(2016)}]{Giovannini:2016whv}%
  \BibitemOpen
  \bibfield  {author} {\bibinfo {author} {\bibfnamefont {M.}~\bibnamefont
  {Giovannini}},\ }\href {\doibase 10.1103/PhysRevD.94.081301} {\bibfield
  {journal} {\bibinfo  {journal} {Phys. Rev. D}\ }\textbf {\bibinfo {volume}
  {94}},\ \bibinfo {pages} {081301} (\bibinfo {year} {2016})},\ \Eprint
  {http://arxiv.org/abs/1606.08205} {arXiv:1606.08205 [hep-th]} \BibitemShut
  {NoStop}%
\bibitem [{\citenamefont {Yamamoto}(2016)}]{Yamamoto:2016xtu}%
  \BibitemOpen
  \bibfield  {author} {\bibinfo {author} {\bibfnamefont {N.}~\bibnamefont
  {Yamamoto}},\ }\href {\doibase 10.1103/PhysRevD.93.125016} {\bibfield
  {journal} {\bibinfo  {journal} {Phys. Rev. D}\ }\textbf {\bibinfo {volume}
  {93}},\ \bibinfo {pages} {125016} (\bibinfo {year} {2016})},\ \Eprint
  {http://arxiv.org/abs/1603.08864} {arXiv:1603.08864 [hep-th]} \BibitemShut
  {NoStop}%
\bibitem [{\citenamefont {Rogachevskii}\ \emph {et~al.}(2017)\citenamefont
  {Rogachevskii}, \citenamefont {Ruchayskiy}, \citenamefont {Boyarsky},
  \citenamefont {Fr\"ohlich}, \citenamefont {Kleeorin}, \citenamefont
  {Brandenburg},\ and\ \citenamefont {Schober}}]{Rogachevskii:2017uyc}%
  \BibitemOpen
  \bibfield  {author} {\bibinfo {author} {\bibfnamefont {I.}~\bibnamefont
  {Rogachevskii}}, \bibinfo {author} {\bibfnamefont {O.}~\bibnamefont
  {Ruchayskiy}}, \bibinfo {author} {\bibfnamefont {A.}~\bibnamefont
  {Boyarsky}}, \bibinfo {author} {\bibfnamefont {J.}~\bibnamefont
  {Fr\"ohlich}}, \bibinfo {author} {\bibfnamefont {N.}~\bibnamefont
  {Kleeorin}}, \bibinfo {author} {\bibfnamefont {A.}~\bibnamefont
  {Brandenburg}}, \ and\ \bibinfo {author} {\bibfnamefont {J.}~\bibnamefont
  {Schober}},\ }\href {\doibase 10.3847/1538-4357/aa886b} {\bibfield  {journal}
  {\bibinfo  {journal} {Astrophys. J.}\ }\textbf {\bibinfo {volume} {846}},\
  \bibinfo {pages} {153} (\bibinfo {year} {2017})},\ \Eprint
  {http://arxiv.org/abs/1705.00378} {arXiv:1705.00378 [physics.plasm-ph]}
  \BibitemShut {NoStop}%
\bibitem [{\citenamefont {Hattori}\ \emph {et~al.}(2019)\citenamefont
  {Hattori}, \citenamefont {Hirono}, \citenamefont {Yee},\ and\ \citenamefont
  {Yin}}]{Hattori:2017usa}%
  \BibitemOpen
  \bibfield  {author} {\bibinfo {author} {\bibfnamefont {K.}~\bibnamefont
  {Hattori}}, \bibinfo {author} {\bibfnamefont {Y.}~\bibnamefont {Hirono}},
  \bibinfo {author} {\bibfnamefont {H.-U.}\ \bibnamefont {Yee}}, \ and\
  \bibinfo {author} {\bibfnamefont {Y.}~\bibnamefont {Yin}},\ }\href {\doibase
  10.1103/PhysRevD.100.065023} {\bibfield  {journal} {\bibinfo  {journal}
  {Phys. Rev. D}\ }\textbf {\bibinfo {volume} {100}},\ \bibinfo {pages}
  {065023} (\bibinfo {year} {2019})},\ \Eprint
  {http://arxiv.org/abs/1711.08450} {arXiv:1711.08450 [hep-th]} \BibitemShut
  {NoStop}%
\bibitem [{\citenamefont {Hattori}\ \emph {et~al.}(2022)\citenamefont
  {Hattori}, \citenamefont {Hongo},\ and\ \citenamefont
  {Huang}}]{Hattori:2022hyo}%
  \BibitemOpen
  \bibfield  {author} {\bibinfo {author} {\bibfnamefont {K.}~\bibnamefont
  {Hattori}}, \bibinfo {author} {\bibfnamefont {M.}~\bibnamefont {Hongo}}, \
  and\ \bibinfo {author} {\bibfnamefont {X.-G.}\ \bibnamefont {Huang}},\ }\href
  {\doibase 10.3390/sym14091851} {\bibfield  {journal} {\bibinfo  {journal}
  {Symmetry}\ }\textbf {\bibinfo {volume} {14}},\ \bibinfo {pages} {1851}
  (\bibinfo {year} {2022})},\ \Eprint {http://arxiv.org/abs/2207.12794}
  {arXiv:2207.12794 [hep-th]} \BibitemShut {NoStop}%
\bibitem [{\citenamefont {Landry}\ and\ \citenamefont {Liu}(pear)}]{LandryLiu}%
  \BibitemOpen
  \bibfield  {author} {\bibinfo {author} {\bibfnamefont {M.~J.}\ \bibnamefont
  {Landry}}\ and\ \bibinfo {author} {\bibfnamefont {H.}~\bibnamefont {Liu}},\
  }\href@noop {} {\  (\bibinfo {year} {to appear})}\BibitemShut {NoStop}%
\bibitem [{\citenamefont {Das}\ \emph {et~al.}(pear)\citenamefont {Das},
  \citenamefont {Iqbal},\ and\ \citenamefont {Poovuttikul}}]{Iqbaletal}%
  \BibitemOpen
  \bibfield  {author} {\bibinfo {author} {\bibfnamefont {A.}~\bibnamefont
  {Das}}, \bibinfo {author} {\bibfnamefont {N.}~\bibnamefont {Iqbal}}, \ and\
  \bibinfo {author} {\bibfnamefont {N.}~\bibnamefont {Poovuttikul}},\
  }\href@noop {} {\  (\bibinfo {year} {to appear})}\BibitemShut {NoStop}%
\bibitem [{\citenamefont {Keldysh}(1964)}]{Keldysh:1964ud}%
  \BibitemOpen
  \bibfield  {author} {\bibinfo {author} {\bibfnamefont {L.}~\bibnamefont
  {Keldysh}},\ }\href@noop {} {\bibfield  {journal} {\bibinfo  {journal}
  {Zh.Eksp.Teor.Fiz.}\ }\textbf {\bibinfo {volume} {47}},\ \bibinfo {pages}
  {1515} (\bibinfo {year} {1964})}\BibitemShut {NoStop}%
\bibitem [{\citenamefont {Crossley}\ \emph {et~al.}(2017)\citenamefont
  {Crossley}, \citenamefont {Glorioso},\ and\ \citenamefont
  {Liu}}]{Crossley:2015evo}%
  \BibitemOpen
  \bibfield  {author} {\bibinfo {author} {\bibfnamefont {M.}~\bibnamefont
  {Crossley}}, \bibinfo {author} {\bibfnamefont {P.}~\bibnamefont {Glorioso}},
  \ and\ \bibinfo {author} {\bibfnamefont {H.}~\bibnamefont {Liu}},\ }\href
  {\doibase 10.1007/JHEP09(2017)095} {\bibfield  {journal} {\bibinfo  {journal}
  {JHEP}\ }\textbf {\bibinfo {volume} {09}},\ \bibinfo {pages} {095} (\bibinfo
  {year} {2017})},\ \Eprint {http://arxiv.org/abs/1511.03646} {arXiv:1511.03646
  [hep-th]} \BibitemShut {NoStop}%
\bibitem [{\citenamefont {Glorioso}\ \emph {et~al.}(2017)\citenamefont
  {Glorioso}, \citenamefont {Crossley},\ and\ \citenamefont
  {Liu}}]{Glorioso:2017fpd}%
  \BibitemOpen
  \bibfield  {author} {\bibinfo {author} {\bibfnamefont {P.}~\bibnamefont
  {Glorioso}}, \bibinfo {author} {\bibfnamefont {M.}~\bibnamefont {Crossley}},
  \ and\ \bibinfo {author} {\bibfnamefont {H.}~\bibnamefont {Liu}},\ }\href
  {\doibase 10.1007/JHEP09(2017)096} {\bibfield  {journal} {\bibinfo  {journal}
  {JHEP}\ }\textbf {\bibinfo {volume} {09}},\ \bibinfo {pages} {096} (\bibinfo
  {year} {2017})},\ \Eprint {http://arxiv.org/abs/1701.07817} {arXiv:1701.07817
  [hep-th]} \BibitemShut {NoStop}%
\bibitem [{\citenamefont {Glorioso}\ and\ \citenamefont
  {Liu}(2018)}]{Glorioso:2018wxw}%
  \BibitemOpen
  \bibfield  {author} {\bibinfo {author} {\bibfnamefont {P.}~\bibnamefont
  {Glorioso}}\ and\ \bibinfo {author} {\bibfnamefont {H.}~\bibnamefont {Liu}},\
  }\href@noop {} {\  (\bibinfo {year} {2018})},\ \Eprint
  {http://arxiv.org/abs/1805.09331} {arXiv:1805.09331 [hep-th]} \BibitemShut
  {NoStop}%
\bibitem [{\citenamefont {Glorioso}\ and\ \citenamefont
  {Liu}(2016)}]{Glorioso:2016gsa}%
  \BibitemOpen
  \bibfield  {author} {\bibinfo {author} {\bibfnamefont {P.}~\bibnamefont
  {Glorioso}}\ and\ \bibinfo {author} {\bibfnamefont {H.}~\bibnamefont {Liu}},\
  }\href@noop {} {\  (\bibinfo {year} {2016})},\ \Eprint
  {http://arxiv.org/abs/1612.07705} {arXiv:1612.07705 [hep-th]} \BibitemShut
  {NoStop}%
\bibitem [{\citenamefont {Kovtun}(2016)}]{Kovtun:2016lfw}%
  \BibitemOpen
  \bibfield  {author} {\bibinfo {author} {\bibfnamefont {P.}~\bibnamefont
  {Kovtun}},\ }\href {\doibase 10.1007/JHEP07(2016)028} {\bibfield  {journal}
  {\bibinfo  {journal} {JHEP}\ }\textbf {\bibinfo {volume} {07}},\ \bibinfo
  {pages} {028} (\bibinfo {year} {2016})},\ \Eprint
  {http://arxiv.org/abs/1606.01226} {arXiv:1606.01226 [hep-th]} \BibitemShut
  {NoStop}%
\bibitem [{\citenamefont {Carroll}\ \emph {et~al.}(1990)\citenamefont
  {Carroll}, \citenamefont {Field},\ and\ \citenamefont
  {Jackiw}}]{PhysRevD.41.1231}%
  \BibitemOpen
  \bibfield  {author} {\bibinfo {author} {\bibfnamefont {S.~M.}\ \bibnamefont
  {Carroll}}, \bibinfo {author} {\bibfnamefont {G.~B.}\ \bibnamefont {Field}},
  \ and\ \bibinfo {author} {\bibfnamefont {R.}~\bibnamefont {Jackiw}},\ }\href
  {\doibase 10.1103/PhysRevD.41.1231} {\bibfield  {journal} {\bibinfo
  {journal} {Phys. Rev. D}\ }\textbf {\bibinfo {volume} {41}},\ \bibinfo
  {pages} {1231} (\bibinfo {year} {1990})}\BibitemShut {NoStop}%
\bibitem [{\citenamefont {Hernandez}\ and\ \citenamefont
  {Kovtun}(2017)}]{Hernandez:2017mch}%
  \BibitemOpen
  \bibfield  {author} {\bibinfo {author} {\bibfnamefont {J.}~\bibnamefont
  {Hernandez}}\ and\ \bibinfo {author} {\bibfnamefont {P.}~\bibnamefont
  {Kovtun}},\ }\href {\doibase 10.1007/JHEP05(2017)001} {\bibfield  {journal}
  {\bibinfo  {journal} {JHEP}\ }\textbf {\bibinfo {volume} {05}},\ \bibinfo
  {pages} {001} (\bibinfo {year} {2017})},\ \Eprint
  {http://arxiv.org/abs/1703.08757} {arXiv:1703.08757 [hep-th]} \BibitemShut
  {NoStop}%
\bibitem [{\citenamefont {Grozdanov}\ and\ \citenamefont
  {Poovuttikul}(2019)}]{Grozdanov:2017kyl}%
  \BibitemOpen
  \bibfield  {author} {\bibinfo {author} {\bibfnamefont {S.}~\bibnamefont
  {Grozdanov}}\ and\ \bibinfo {author} {\bibfnamefont {N.}~\bibnamefont
  {Poovuttikul}},\ }\href {\doibase 10.1007/JHEP04(2019)141} {\bibfield
  {journal} {\bibinfo  {journal} {JHEP}\ }\textbf {\bibinfo {volume} {04}},\
  \bibinfo {pages} {141} (\bibinfo {year} {2019})},\ \Eprint
  {http://arxiv.org/abs/1707.04182} {arXiv:1707.04182 [hep-th]} \BibitemShut
  {NoStop}%
\bibitem [{\citenamefont {Glorioso}\ and\ \citenamefont
  {Son}(2018)}]{Glorioso:2018kcp}%
  \BibitemOpen
  \bibfield  {author} {\bibinfo {author} {\bibfnamefont {P.}~\bibnamefont
  {Glorioso}}\ and\ \bibinfo {author} {\bibfnamefont {D.~T.}\ \bibnamefont
  {Son}},\ }\href@noop {} {\  (\bibinfo {year} {2018})},\ \Eprint
  {http://arxiv.org/abs/1811.04879} {arXiv:1811.04879 [hep-th]} \BibitemShut
  {NoStop}%
\bibitem [{\citenamefont {Armas}\ and\ \citenamefont
  {Jain}(2020)}]{Armas:2018zbe}%
  \BibitemOpen
  \bibfield  {author} {\bibinfo {author} {\bibfnamefont {J.}~\bibnamefont
  {Armas}}\ and\ \bibinfo {author} {\bibfnamefont {A.}~\bibnamefont {Jain}},\
  }\href {\doibase 10.1007/JHEP01(2020)041} {\bibfield  {journal} {\bibinfo
  {journal} {JHEP}\ }\textbf {\bibinfo {volume} {01}},\ \bibinfo {pages} {041}
  (\bibinfo {year} {2020})},\ \Eprint {http://arxiv.org/abs/1811.04913}
  {arXiv:1811.04913 [hep-th]} \BibitemShut {NoStop}%
\bibitem [{\citenamefont {Gralla}\ and\ \citenamefont
  {Iqbal}(2019)}]{Gralla:2018kif}%
  \BibitemOpen
  \bibfield  {author} {\bibinfo {author} {\bibfnamefont {S.~E.}\ \bibnamefont
  {Gralla}}\ and\ \bibinfo {author} {\bibfnamefont {N.}~\bibnamefont {Iqbal}},\
  }\href {\doibase 10.1103/PhysRevD.99.105004} {\bibfield  {journal} {\bibinfo
  {journal} {Phys. Rev. D}\ }\textbf {\bibinfo {volume} {99}},\ \bibinfo
  {pages} {105004} (\bibinfo {year} {2019})},\ \Eprint
  {http://arxiv.org/abs/1811.07438} {arXiv:1811.07438 [hep-th]} \BibitemShut
  {NoStop}%
\bibitem [{\citenamefont {Benenowski}\ and\ \citenamefont
  {Poovuttikul}(2019)}]{Benenowski:2019ule}%
  \BibitemOpen
  \bibfield  {author} {\bibinfo {author} {\bibfnamefont {B.}~\bibnamefont
  {Benenowski}}\ and\ \bibinfo {author} {\bibfnamefont {N.}~\bibnamefont
  {Poovuttikul}},\ }\href@noop {} {\  (\bibinfo {year} {2019})},\ \Eprint
  {http://arxiv.org/abs/1911.05554} {arXiv:1911.05554 [hep-th]} \BibitemShut
  {NoStop}%
\bibitem [{\citenamefont {Landry}(2021)}]{Landry:2021kko}%
  \BibitemOpen
  \bibfield  {author} {\bibinfo {author} {\bibfnamefont {M.~J.}\ \bibnamefont
  {Landry}},\ }\href@noop {} {\  (\bibinfo {year} {2021})},\ \Eprint
  {http://arxiv.org/abs/2101.02210} {arXiv:2101.02210 [hep-th]} \BibitemShut
  {NoStop}%
\bibitem [{\citenamefont {Armas}\ and\ \citenamefont
  {Camilloni}(2022)}]{Armas:2022wvb}%
  \BibitemOpen
  \bibfield  {author} {\bibinfo {author} {\bibfnamefont {J.}~\bibnamefont
  {Armas}}\ and\ \bibinfo {author} {\bibfnamefont {F.}~\bibnamefont
  {Camilloni}},\ }\href@noop {} {\  (\bibinfo {year} {2022})},\ \Eprint
  {http://arxiv.org/abs/2201.06847} {arXiv:2201.06847 [hep-th]} \BibitemShut
  {NoStop}%
\bibitem [{\citenamefont {{Goldreich}}\ and\ \citenamefont
  {{Reisenegger}}(1992)}]{1992ApJ...395..250G}%
  \BibitemOpen
  \bibfield  {author} {\bibinfo {author} {\bibfnamefont {P.}~\bibnamefont
  {{Goldreich}}}\ and\ \bibinfo {author} {\bibfnamefont {A.}~\bibnamefont
  {{Reisenegger}}},\ }\href {\doibase 10.1086/171646} {\bibfield  {journal}
  {\bibinfo  {journal} {\apj}\ }\textbf {\bibinfo {volume} {395}},\ \bibinfo
  {pages} {250} (\bibinfo {year} {1992})}\BibitemShut {NoStop}%
\bibitem [{\citenamefont {Glorioso}\ \emph {et~al.}(2019)\citenamefont
  {Glorioso}, \citenamefont {Liu},\ and\ \citenamefont
  {Rajagopal}}]{Glorioso:2017lcn}%
  \BibitemOpen
  \bibfield  {author} {\bibinfo {author} {\bibfnamefont {P.}~\bibnamefont
  {Glorioso}}, \bibinfo {author} {\bibfnamefont {H.}~\bibnamefont {Liu}}, \
  and\ \bibinfo {author} {\bibfnamefont {S.}~\bibnamefont {Rajagopal}},\ }\href
  {\doibase 10.1007/JHEP01(2019)043} {\bibfield  {journal} {\bibinfo  {journal}
  {JHEP}\ }\textbf {\bibinfo {volume} {01}},\ \bibinfo {pages} {043} (\bibinfo
  {year} {2019})},\ \Eprint {http://arxiv.org/abs/1710.03768} {arXiv:1710.03768
  [hep-th]} \BibitemShut {NoStop}%
\bibitem [{\citenamefont {Son}\ and\ \citenamefont
  {Zhitnitsky}(2004)}]{Son:2004tq}%
  \BibitemOpen
  \bibfield  {author} {\bibinfo {author} {\bibfnamefont {D.~T.}\ \bibnamefont
  {Son}}\ and\ \bibinfo {author} {\bibfnamefont {A.~R.}\ \bibnamefont
  {Zhitnitsky}},\ }\href {\doibase 10.1103/PhysRevD.70.074018} {\bibfield
  {journal} {\bibinfo  {journal} {Phys. Rev. D}\ }\textbf {\bibinfo {volume}
  {70}},\ \bibinfo {pages} {074018} (\bibinfo {year} {2004})},\ \Eprint
  {http://arxiv.org/abs/hep-ph/0405216} {arXiv:hep-ph/0405216} \BibitemShut
  {NoStop}%
\bibitem [{\citenamefont {Metlitski}\ and\ \citenamefont
  {Zhitnitsky}(2005)}]{PhysRevD.72.045011}%
  \BibitemOpen
  \bibfield  {author} {\bibinfo {author} {\bibfnamefont {M.~A.}\ \bibnamefont
  {Metlitski}}\ and\ \bibinfo {author} {\bibfnamefont {A.~R.}\ \bibnamefont
  {Zhitnitsky}},\ }\href {\doibase 10.1103/PhysRevD.72.045011} {\bibfield
  {journal} {\bibinfo  {journal} {Phys. Rev. D}\ }\textbf {\bibinfo {volume}
  {72}},\ \bibinfo {pages} {045011} (\bibinfo {year} {2005})}\BibitemShut
  {NoStop}%
\bibitem [{\citenamefont {Huang}\ and\ \citenamefont
  {Liao}(2013)}]{PhysRevLett.110.232302}%
  \BibitemOpen
  \bibfield  {author} {\bibinfo {author} {\bibfnamefont {X.-G.}\ \bibnamefont
  {Huang}}\ and\ \bibinfo {author} {\bibfnamefont {J.}~\bibnamefont {Liao}},\
  }\href {\doibase 10.1103/PhysRevLett.110.232302} {\bibfield  {journal}
  {\bibinfo  {journal} {Phys. Rev. Lett.}\ }\textbf {\bibinfo {volume} {110}},\
  \bibinfo {pages} {232302} (\bibinfo {year} {2013})}\BibitemShut {NoStop}%
\bibitem [{\citenamefont {Jiang}\ \emph {et~al.}(2015)\citenamefont {Jiang},
  \citenamefont {Huang},\ and\ \citenamefont {Liao}}]{PhysRevD.91.045001}%
  \BibitemOpen
  \bibfield  {author} {\bibinfo {author} {\bibfnamefont {Y.}~\bibnamefont
  {Jiang}}, \bibinfo {author} {\bibfnamefont {X.-G.}\ \bibnamefont {Huang}}, \
  and\ \bibinfo {author} {\bibfnamefont {J.}~\bibnamefont {Liao}},\ }\href
  {\doibase 10.1103/PhysRevD.91.045001} {\bibfield  {journal} {\bibinfo
  {journal} {Phys. Rev. D}\ }\textbf {\bibinfo {volume} {91}},\ \bibinfo
  {pages} {045001} (\bibinfo {year} {2015})}\BibitemShut {NoStop}%
\bibitem [{\citenamefont {Kharzeev}\ and\ \citenamefont
  {Yee}(2011)}]{Kharzeev:2010gd}%
  \BibitemOpen
  \bibfield  {author} {\bibinfo {author} {\bibfnamefont {D.~E.}\ \bibnamefont
  {Kharzeev}}\ and\ \bibinfo {author} {\bibfnamefont {H.-U.}\ \bibnamefont
  {Yee}},\ }\href {\doibase 10.1103/PhysRevD.83.085007} {\bibfield  {journal}
  {\bibinfo  {journal} {Phys. Rev. D}\ }\textbf {\bibinfo {volume} {83}},\
  \bibinfo {pages} {085007} (\bibinfo {year} {2011})},\ \Eprint
  {http://arxiv.org/abs/1012.6026} {arXiv:1012.6026 [hep-th]} \BibitemShut
  {NoStop}%
\bibitem [{\citenamefont {Vilenkin}(1980)}]{PhysRevD.22.3080}%
  \BibitemOpen
  \bibfield  {author} {\bibinfo {author} {\bibfnamefont {A.}~\bibnamefont
  {Vilenkin}},\ }\href {\doibase 10.1103/PhysRevD.22.3080} {\bibfield
  {journal} {\bibinfo  {journal} {Phys. Rev. D}\ }\textbf {\bibinfo {volume}
  {22}},\ \bibinfo {pages} {3080} (\bibinfo {year} {1980})}\BibitemShut
  {NoStop}%
\bibitem [{\citenamefont {Kharzeev}\ \emph {et~al.}(2008)\citenamefont
  {Kharzeev}, \citenamefont {McLerran},\ and\ \citenamefont
  {Warringa}}]{Kharzeev:2007jp}%
  \BibitemOpen
  \bibfield  {author} {\bibinfo {author} {\bibfnamefont {D.~E.}\ \bibnamefont
  {Kharzeev}}, \bibinfo {author} {\bibfnamefont {L.~D.}\ \bibnamefont
  {McLerran}}, \ and\ \bibinfo {author} {\bibfnamefont {H.~J.}\ \bibnamefont
  {Warringa}},\ }\href {\doibase 10.1016/j.nuclphysa.2008.02.298} {\bibfield
  {journal} {\bibinfo  {journal} {Nucl. Phys. A}\ }\textbf {\bibinfo {volume}
  {803}},\ \bibinfo {pages} {227} (\bibinfo {year} {2008})},\ \Eprint
  {http://arxiv.org/abs/0711.0950} {arXiv:0711.0950 [hep-ph]} \BibitemShut
  {NoStop}%
\bibitem [{\citenamefont {Fukushima}\ \emph {et~al.}(2008)\citenamefont
  {Fukushima}, \citenamefont {Kharzeev},\ and\ \citenamefont
  {Warringa}}]{Fukushima:2008xe}%
  \BibitemOpen
  \bibfield  {author} {\bibinfo {author} {\bibfnamefont {K.}~\bibnamefont
  {Fukushima}}, \bibinfo {author} {\bibfnamefont {D.~E.}\ \bibnamefont
  {Kharzeev}}, \ and\ \bibinfo {author} {\bibfnamefont {H.~J.}\ \bibnamefont
  {Warringa}},\ }\href {\doibase 10.1103/PhysRevD.78.074033} {\bibfield
  {journal} {\bibinfo  {journal} {Phys. Rev. D}\ }\textbf {\bibinfo {volume}
  {78}},\ \bibinfo {pages} {074033} (\bibinfo {year} {2008})},\ \Eprint
  {http://arxiv.org/abs/0808.3382} {arXiv:0808.3382 [hep-ph]} \BibitemShut
  {NoStop}%
\bibitem [{\citenamefont {Kharzeev}(2014)}]{Kharzeev:2013ffa}%
  \BibitemOpen
  \bibfield  {author} {\bibinfo {author} {\bibfnamefont {D.~E.}\ \bibnamefont
  {Kharzeev}},\ }\href {\doibase 10.1016/j.ppnp.2014.01.002} {\bibfield
  {journal} {\bibinfo  {journal} {Prog. Part. Nucl. Phys.}\ }\textbf {\bibinfo
  {volume} {75}},\ \bibinfo {pages} {133} (\bibinfo {year} {2014})},\ \Eprint
  {http://arxiv.org/abs/1312.3348} {arXiv:1312.3348 [hep-ph]} \BibitemShut
  {NoStop}%
\bibitem [{\citenamefont {Akamatsu}\ and\ \citenamefont
  {Yamamoto}(2013)}]{Akamatsu:2013pjd}%
  \BibitemOpen
  \bibfield  {author} {\bibinfo {author} {\bibfnamefont {Y.}~\bibnamefont
  {Akamatsu}}\ and\ \bibinfo {author} {\bibfnamefont {N.}~\bibnamefont
  {Yamamoto}},\ }\href {\doibase 10.1103/PhysRevLett.111.052002} {\bibfield
  {journal} {\bibinfo  {journal} {Phys. Rev. Lett.}\ }\textbf {\bibinfo
  {volume} {111}},\ \bibinfo {pages} {052002} (\bibinfo {year} {2013})},\
  \Eprint {http://arxiv.org/abs/1302.2125} {arXiv:1302.2125 [nucl-th]}
  \BibitemShut {NoStop}%
\bibitem [{\citenamefont {Kaplan}\ \emph {et~al.}(2017)\citenamefont {Kaplan},
  \citenamefont {Reddy},\ and\ \citenamefont {Sen}}]{Kaplan:2016drz}%
  \BibitemOpen
  \bibfield  {author} {\bibinfo {author} {\bibfnamefont {D.~B.}\ \bibnamefont
  {Kaplan}}, \bibinfo {author} {\bibfnamefont {S.}~\bibnamefont {Reddy}}, \
  and\ \bibinfo {author} {\bibfnamefont {S.}~\bibnamefont {Sen}},\ }\href
  {\doibase 10.1103/PhysRevD.96.016008} {\bibfield  {journal} {\bibinfo
  {journal} {Phys. Rev. D}\ }\textbf {\bibinfo {volume} {96}},\ \bibinfo
  {pages} {016008} (\bibinfo {year} {2017})},\ \Eprint
  {http://arxiv.org/abs/1612.00032} {arXiv:1612.00032 [hep-ph]} \BibitemShut
  {NoStop}%
\end{thebibliography}%

\end{document}